\documentclass[12pt]{JHEP3}

\usepackage{amsmath,amssymb,amscd,amsfonts,xspace,mathrsfs,amsthm,dsfont}
\usepackage{bbm}
\usepackage{tikz}
\usepackage{steinmetz}
\usepackage{braket}
\usepackage{caption}
\usepackage{color}
\let\normalcolor\relax
    \definecolor{darkgreen}{rgb}{0,0.5,0}
    \definecolor{darkred}{rgb}{0.5,0,0}
    \definecolor{darkblue}{rgb}{0,0,0.6}
    \definecolor{purple}{rgb}{0.4,.2,0.7}

\usepackage[mathcal]{eucal}
\usepackage{float}
\usepackage{mathtools}
\usepackage{pdfsync}
\usepackage{slashed}
\usepackage[normalem]{ulem}
\usepackage{upgreek}
\usepackage{url}
\usepackage{diagbox}

\usepackage{import}
\usepackage{pdfpages}
\usepackage{lmodern}
\usepackage{mathtools}

\usepackage[T1]{fontenc} % if needed
\gdef\@fpheader{}
\usepackage[ragged]{footmisc}
\newcommand{\ben}{\begin{eqnarray}\displaystyle}
\newcommand{\een}{\end{eqnarray}}
\def\bea{\begin{eqnarray}}
\def\eea{\end{eqnarray}}
\def\be{\begin{equation}}
\def\ee{\end{equation}}
\def\ba{\begin{align}}
\def\ea{\end{align}}
\def\bse{\begin{subequations}}
\def\ese{\end{subequations}}

\numberwithin{equation}{section}

\def\Sym{\,{\rm Sym}\, }

\def\Im{\,{\rm Im}\,}
\def\Re{\,{\rm Re}\,}

\def\Li{{\rm Li}}
\newcommand{\p}{\partial}

\renewcommand{\tilde}{\widetilde}

\def\({\left(}
\def\){\right)}
\def\[{\left[}
\def\]{\right]}
\def\<{\left\langle}
\def\>{\right\rangle}
\def\hf{{1\over 2}}

\newcommand{\cA}{\mathcal{A}}
\newcommand{\cB}{\mathcal{B}}

\newcommand{\cF}{\mathcal{F}}

\newcommand{\cI}{\mathcal{I}}

\newcommand{\cN}{\mathcal{N}}
\newcommand{\cO}{\mathcal{O}}

\newcommand{\cS}{\mathcal{S}}

\newcommand{\cZ}{\mathcal{Z}}

\newcommand{\bfn}{{\boldsymbol n}}

\newcommand{\bfx}{{\boldsymbol x}}

\newcommand{\IR}{\mathds{R}}

\newcommand{\IZ}{\mathds{Z}}

\newcommand{\IN}{\mathds{N}}

\newcommand{\IS}{\mathds{S}}

\def\gsf{\mathsf{g}}

\newcommand{\nn}{\nonumber}

\newcommand{\eps}{\epsilon}

\def\tgsf{\tilde\gsf}

\def\xp{x_{+}}
\def\xm{x_{-}}
\def\xpm{x_{\pm}}

\def\pse{ \psi^{_{E}} }

\def\psep{\pse_+}
\def\psem{\pse_-}
\def\psepm{\pse_\pm}

\def\Ci#1{C^{(#1)}}
\def\Zi#1{Z^{(#1)}}
\def\whZi#1{\widehat Z^{(#1)}}
\def\cIi#1{\cI^{(#1)}}
\def\tcIi#1{\tilde\cI^{(#1)}}
\def\whcIi#1{\widehat\cI^{(#1)}}

\def\Lami#1{\Lambda^{(#1)}}

\def\whS{\widehat S}

\def\chf{\check f}
\def\chzeta{\check\zeta}

\def\Res#1{\mathop{\rm Res}\limits_{#1}}

\title{Multi-instantons in 2d string theory}

\author{Sergei Alexandrov$^1$ and Rishabh Kaushik$^2$
\\
$^1$ {\it
Laboratoire Charles Coulomb (L2C), Universit\'e de Montpellier,
CNRS, F-34095, Montpellier, France}\\
\\
$^2$ {\it International Center for Theoretical Sciences(ICTS-TIFR),
	Shivakote, Hesaraghatta Hobli, Bengaluru North 560089, India}

\vspace*{2mm} {\tt e-mail:
\email{sergey.alexandrov@umontpellier.fr},
\email{rishabh.kaushik@icts.res.in}
}

\vspace*{-3mm}
}

\abstract{
Instanton contributions in 2d string theory are known to include subtle numerical factors $\zeta_n$
closely related to a contour prescription in multi-instanton string amplitudes.
Both ingredients appear to be ambiguous due to a degeneracy between $(1,n)$-ZZ instantons 
and $n$ (1,1)-ZZ instantons in the linear dilaton background. We resolve this ambiguity 
using insights from the dual matrix quantum mechanics where the multipliers $\zeta_n$
can be derived from an integral representation of the scattering phase and 
follow from the median resummation prescribed by resurgence theory.
We evaluate multi-instanton string amplitudes in the theory compactified on a circle of finite radius
for arbitrary number of instantons and show that they reproduce the matrix model predictions
provided the Lorentzian contour prescription is used for their evaluation.
We also show that the non-perturbative free energy matches the structure of the D-instanton induced 
string field theory effective action, which suggests the vanishing of contributions from worldsheet topologies 
of negative Euler number.
}

\begin{document}

\baselineskip 15.pt
\setlength{\parskip}{0.13cm}

\section{Introduction}

Recently, due to the works of Ashoke Sen \cite{Sen:2020cef,Sen:2020eck,Sen:2021qdk},
there has been a significant progress in the direct evaluation of D-instanton string amplitudes.
Using insights from string field theory, he understood how to properly regularize 
the divergences appearing in the annulus amplitudes with D-brane boundary conditions evaluated in the naive CFT approach. 
This has allowed not only to reproduce (see \cite{Sen:2020eck,Sen:2021qdk,Sen:2021tpp,Sen:2021jbr,Alexandrov:2021shf,Alexandrov:2021dyl,Eniceicu:2022nay,Eniceicu:2022dru,Chakravarty:2022cgj,Eniceicu:2022xvk,Alexandrov:2023fvb,Chakrabhavi:2024szk,Kaushik:2025neu}) 
the results in critical and non-critical string theories 
obtained by other means, such as dualities and matrix model techniques,
but also to get genuinely new results about normalization of D-instanton effects \cite{Alexandrov:2022mmy}. 

While in critical string theory the recipe of \cite{Sen:2020cef,Sen:2020eck,Sen:2021qdk} gives the complete normalization of 
D-instanton amplitudes, in non-critical strings the normalization turns out to involve an additional numerical factor $\zeta$,
which has been attributed in \cite{Sen:2020ruy,Sen:2021qdk} to the fact that the actual contour of integration 
over some string theoretical degrees of freedom 
is not identical to the steepest descent contour associated with the D-instanton.
In the case of the leading one-instanton effect, this factor is well-known to be equal to 1/2.
However, beyond this case, it remained undetermined.

In this paper, we address this problem for two-dimensional string theory
(see \cite{Klebanov:1991qa,Ginsparg:1993is,Jevicki:1993qn,Polchinski:1994mb} for reviews).
It has a dual description in terms of the double scaled matrix quantum mechanics (MQM) \cite{Brezin:1989ss,Ginsparg:1990as,Gross:1990ay},
which is exactly solvable. In particular, using this description, the full non-perturbative partition
function for the theory in the linear dilaton background (and placed at finite temperature) has been found long ago
\cite{Alexandrov:2004cg}, generalizing the earlier perturbative formula \cite{Klebanov:1991qa}.
So, one could think that this result should be enough to predict all relevant factors $\zeta$.
However, it is not because of a degeneracy inherent to this background.

To explain the problem, let us recall that in Liouville theory, which represents the main non-trivial part of
the worldsheet description of non-critical strings, there is a two-parameter family of Dirichlet 
boundary conditions \cite{Zamolodchikov:2001ah}, giving rise to the so-called $(m,n)$ ZZ-branes.
In \cite{Alexandrov:2003un}, it was found (and rediscovered 16 years later in \cite{Balthazar:2019ypi}) 
that 2d string theory includes only the one-parameter subset of $(1,n)$ ZZ-branes. 
In principle, each of the instantons generated by these branes can have its own normalization factor $\zeta_n$. 
On the other hand, the normalization factors of multi-instantons have been argued in \cite{Balthazar:2019ypi,Kaushik:2025neu} 
to be given by the product of the normalization factors of the individual instantons.
Thus, we have to determine a set of numbers $\zeta_n$ assigned to $(1,n)$ ZZ-instantons.
The problem is that in the linear dilaton background the instanton actions for $(1,n)$ ZZ-instanton 
and $n$ $(1,1)$ ZZ-instantons are equal. This leads to a degeneracy 
that does not allow to distinguish their contributions and makes it impossible to read $\zeta_n$
from the formula for the partition function.

A way out could be to consider a background with a non-trivial tachyon potential as in
\cite{Alexandrov:2002fh,Alexandrov:2004cg,Alexandrov:2023fvb}, 
which removes the degeneracy. This is precisely how the set of $(1,n)$ ZZ-branes has been identified in \cite{Alexandrov:2003un}.
But the problem is that the full non-perturbative partition function in such backgrounds is unknown yet,
and the matrix model evaluation of the instantons is expected to involve the same factors $\zeta_n$ 
that appear in the normalization of string amplitudes.

Instead, we combine the latter observation with another one following from the analysis of \cite{Alexandrov:2023fvb}
that ZZ-instantons are in the one-to-one correspondence with the saddle points of 
an integral representation of the scattering phase in the chiral formalism of MQM \cite{Alexandrov:2002fh}
(see \cite{Alexandrov:2003ut} for a review).
In the linear dilaton background, this scattering phase is well-known and proportional to the Gamma function.
Thus, the factors $\zeta_n$ can be simply obtained by comparing the non-perturbative part of the Gamma function 
with the saddle point contributions of its integral representation.  
The result is given by 
\be  
\zeta_n=\frac{(2n)!}{2^{2n}(n!)^2}=\frac{(2n-1)!!}{(2n)!!}\, .
\label{reszetan}
\ee 
Although it remains puzzling how such rational numbers can arise from the geometry of integration contours, 
we find that they perfectly match the so-called median resummation prescription, 
well-known in resurgence theory \cite{Marino:2008ya,Aniceto:2013fka}, applied to the Gamma function.

Once $\zeta_n$ are known, one can express the non-perturbative part of the partition function 
as an expansion in their powers.
Then the coefficient of a monomial $\zeta_{n_1}^{N_1}\cdots \zeta_{n_k}^{N_k}$ gives a prediction 
for the instanton amplitude generated by $N_1$ $(1,n_1)$ ZZ-instantons, $N_2$ $(1,n_2)$ ZZ-instantons, etc. 
In the second part of the paper, we verify these predictions by explicitly evaluating the relevant multi-instanton string amplitudes
for arbitrary number of instantons.

At this point, it should be mentioned that the multi-instanton string amplitudes involve integrals over open string zero modes, 
related to the positions of D-instantons in the Euclidean time direction, with integrands having poles
at a discrete set of points. Therefore, these integrals require a contour prescription to produce finite results.
Two natural prescriptions have been proposed in the literature: Lorentzian \cite{Balthazar:2019ypi} and unitary \cite{Sen:2020ruy}.
Whereas in unitary theories, such as critical string theory, the unitary prescription must be used, in non-unitary models,
such as 2d string theory, both or even more options could be possible \cite{Sen:2020ruy}.
As we demonstrate, the result \eqref{reszetan} uniquely chooses the Lorentzian prescription.
In particular, the amplitudes we calculate in the zero temperature limit reproduce the normalization factors 
found in \cite{Balthazar:2019ypi} from the S-matrix analysis.

These results allow us to rewrite the non-perturbative free energy in terms of string amplitudes.
It turns out to have exactly the same structure as the string field theory (SFT) effective action induced by D-instantons
\cite{Sen:2024zqr}. Furthermore, to match the two quantities, one should neglect contributions of all 
worldsheet topologies of negative Euler number. Since the expression for the free energy 
given by MQM is supposed to be exact, this suggests a conjecture
that all such contributions should vanish in the linear dilaton background.

The organization of the paper is as follows. 
In the next section, we review the instanton calculus in 2d string theory and formulate the problem.
In section \ref{sec-MQM}, we analyze it in the dual MQM description and propose a solution
from an integral representation of the scattering phase. We justify it from resurgence theory
and produce a set of predictions for D-instanton string amplitudes.
In section \ref{sec-string}, we verify these predictions by explicitly evaluating the amplitudes, 
establish a connection with the SFT effective action, and obtain the zero temperature limit.
In section \ref{sec-disc}, we present our conclusions. 
In appendix \ref{ap-signs}, we discuss certain subtleties appearing in the chiral formulation of MQM 
that have been missed before and lead to sign flips in non-perturbative contributions to various quantities.
Finally, in appendix \ref{ap-unitary}, we derive the values of $\zeta_n$ that would follow from the unitary prescription
and discuss their compatibility with MQM.

\section{Instanton contributions in compactified 2d string theory}
\label{sec-instanton-string}

In this section, we briefly recall the general form of multi-instanton string amplitudes.
We will be interested in the particular case of 2d string theory compactified on a Euclidean circle of radius $R$,
which is equivalent to putting the theory at finite temperature $1/(2\pi R)$.
In the linear dilaton background, the worldsheet description of the theory is given by a direct product
of Liouville theory at central charge $c=25$, a compactified free boson with $c=1$ and a ghost CFT with $c=-26$.
The boundary conditions giving rise to instanton effects are obtained by imposing the $(1,n)$ ZZ-boundary conditions 
\cite{Zamolodchikov:2001ah} on the Liouville field and either Dirichlet or Neumann boundary conditions 
on the free boson, which are parametrized by the position on the circle or its dual, 
respectively.\footnote{At the self-dual radius $R=1$, instead of the two possibilities parametrized by $S^1$, 
	the boundary conditions on the free boson are parametrized by an $SU(2)$ element 
	\cite{Recknagel:1998ih,Gaberdiel:2001xm,Recknagel:2013uja}. 
	We refer to \cite{Kaushik:2025neu} for the analysis of D-instanton contributions in this case.}
Since the latter are T-dual to the former, and the annulus amplitude with mixed boundary conditions 
(Dirichlet on one boundary and Neumann on the other) vanishes \cite{Kaushik:2025neu}, we will concentrate 
mostly on the Dirichlet conditions.

The rules of the D-instanton calculus have been established in \cite{Sen:2020cef}.
Applied to the partition function in 2d string theory, the quantity we are interested in here, they dictate that
\be 
Z(g_s)=Z_{\rm pert}(g_s) Z_{\rm np,D}(g_s)Z_{\rm np,N}(g_s),
\label{partfunZ}
\ee 
where the first factor $Z_{\rm pert}$ is the perturbative contribution given by closed string loop diagrams, 
while the last two factors are the non-perturbative contributions given by 
exponentiated connected open string diagrams with all boundaries ending on ZZ-branes with Dirichlet and Neumann 
boundary conditions, respectively, along the compactified direction. 
Furthermore, we should sum over all possible instanton boundary conditions.
This leads to the following expression for the first non-perturbative factor
\be 
Z_{\rm np,D}(g_s)=1+\sum_{N=1}^\infty\frac{1}{N!} \prod_{i=1}^N 
\[\sum_{n_i=1}^\infty \int\limits_0^{2\pi R} d x_i  \] \cN(\bfn,\bfx)
\exp\(-\sum_{i=1}^N S_{n_i}(x_i)+\cO(g_s)\),
\label{Znp}
\ee 
where $N$ corresponds to the number of instantons, 
the variables $(n_i,x_i)\in\IN\times S^1$ parametrize the boundary conditions of the $i$-th instanton, 
$\bfn$ and $\bfx$ denote the collections of these parameters: $\bfn=(n_1,\dots, n_N)$, $\bfx=(x_1,\dots, x_N)$,
$S_{n}(x)$ is the instanton action given by minus the disk amplitude with the boundary conditions labeled by $(n,x)$
and scaling as $g_s^{-1}$,
$\cN(\bfn,\bfx)$ is a normalization factor given by the exponentiated annulus amplitudes which scale as $g_s^0$, 
and we neglected all other amplitudes which scale as positive powers of the string coupling.

In the linear dilaton background, the instanton action $S_n$ is particularly simple: it is independent of $x$ and,
in the proper normalization, is given by
\be
S_{n}(x)=n/g_s.
\label{instact}
\ee 
However, this simplicity is the reason for a degeneracy: all instanton configurations with a fixed sum 
of labels $n_i$ have the same strength. As a result, all such contributions are mixed together 
and it is hard to distinguish them from each other.

The normalization factor can formally be written as 
\be 
\cN(\bfn,\bfx)=(2\pi g_s)^{-N/2} \(\prod_{i=1}^N \zeta_{n_i}\) 
\exp\(\sum_{i=1}^N \int\limits_0^\infty \frac{dt}{2t}\, \cA_{n_i}(t)
+\sum_{i<j}\int\limits_0^\infty \frac{dt}{t}\, \cA_{n_i,n_j}(t;x_{ij})\),
\label{defN}
\ee 
where $\cA_n(t)$ is the annulus amplitude with both boundaries on the same $(1,n)$ ZZ-instanton,
$\cA_{n,m}(t;x)$ is the annulus amplitude with boundaries on $(1,n)$ and $(1,m)$ ZZ-instantons
separated by distance $x$, and $x_{ij}=x_i-x_j$.
The factor $(2\pi g_s)^{-N/2} $ arises from a difference in the normalization of the open string zero mode and 
the coordinate on the time circle \cite{Sen:2021qdk,Alexandrov:2023fvb}.
Finally, the multipliers $\zeta_n$ are a feature of non-critical string models and 
expected to be determined by how the steepest descent contour associated with the D-instanton 
fits inside the actual integration contour of the string path integral \cite{Sen:2020ruy,Sen:2021qdk}.
While it is known that $\zeta_1=1/2$, we are not aware of any results for $n>1$.
So our first goal will be to find these multipliers.

Before we approach this goal, let us finish defining the normalization factor \eqref{defN}.
This expression is formal because the integrals over the annulus parameter $t$ of 
the annulus amplitudes naively evaluated from the worksheet CFT are divergent 
and have to be regularized as prescribed in \cite{Sen:2020cef,Sen:2021qdk}.
This results in \cite{Alexandrov:2023fvb,Kaushik:2025neu}\footnote{We have added the sign factor $(-1)^{n-1}$ 
compared to \cite{Kaushik:2025neu}. We are free to do this because of the ambiguity in the analytic 
continuation of the tachyonic modes contributing to the annulus amplitude.}
\be 
\exp\(\int_0^\infty \frac{dt}{2t}\, \cA_{n}\)=\frac{(-1)^{n-1} i\sqrt{2\pi  g_s}}{4\pi R\sin(\pi n/R)}
\label{one-an}
\ee 
and \cite{Kaushik:2025neu}
\be 
\exp\(\int_0^\infty \frac{dt}{t}\, \cA_{n,m}(x)\)=g_{n,m}\(\frac{x}{2\pi R}\),
\label{two-an}
\ee 
where we introduced
\be 
g_{n,m}(y)=
\frac{\sin\pi\(y-\frac{|n-m|}{2R}\)\sin\pi\(y+\frac{|n-m|}{2R}\)}
{\sin\pi\(y-\frac{n+m}{2R}\)\sin\pi\(y+\frac{n+m}{2R}\)}\, .
\label{def-gmn}
\ee 
Substituting all these results into \eqref{Znp} and denoting 
\bea  
\cIi{N}_\bfn&=&\int\limits_0^1 \prod_{i=1}^N d y_i \,  \prod_{i<j} g_{n_i,n_j}(y_{ij}),
\label{defIn}
\\
\Zi{1}_n &=&\frac{(-1)^{n} \zeta_{n}}{2i\sin(\pi n/R)}\, e^{-n/g_s},
\label{normfac}
\eea 
one arrives at 
\be 
Z_{\rm np,D}(g_s)=1+\sum_{N=1}^\infty\frac{1}{N!} \prod_{i=1}^N \[
\sum_{n_i=1}^\infty \Zi{1}_{n_i} \]\cIi{N}_\bfn.
\label{Znp-res}
\ee 
A few comments are in order:
\begin{itemize}
	\item 
	For $N=1$, the integral \eqref{defIn} is trivial, $\cIi{1}_n=1$. 
	\item 
	For $N>1$, since $g_{m,n}(y)$ is an even function, $\cIi{N}_\bfn$ does not depend on the order of $n_i$'s in $\bfn$.
	Thus, it is a function of the set $\{n_1,\dots,n_N\}$.
	\item 
	Due to this fact, one can collect the terms in \eqref{Znp-res} corresponding to the same set of $n_i$'s.
	This gives rise to an additional symmetry factor $\frac{N!}{N_1!\cdots N_k!} $ where $N_j$ are the numbers of branes
	with the same ZZ label.
	\item 
	Importantly, the function $g_{n,m}(y)$ has poles on the real axis at $y=\pm\frac{n+m}{2R}+2\pi k$, $k\in\IZ$.
	Therefore, to produce a finite result, the integral \eqref{defIn} should be supplemented by a contour prescription.
	It has been discussed in \cite{Balthazar:2019ypi,Sen:2020ruy} and we will see that it is the Lorentzian prescription of 
	\cite{Balthazar:2019ypi} that will be favored by our results.
\end{itemize}

Finally, the second non-perturbative factor in \eqref{partfunZ}, $Z_{\rm np,N}$, has exactly the same representation as \eqref{Znp}
except that the disk and annulus amplitudes are evaluated with the Neumann boundary conditions
on the free boson and $x_i$ run over the dual circle of radius $1/R$.
As a result, it can be written in the same form as \eqref{Znp-res} where $R$ and $g_s$ are replaced by $1/R$ and $g_s/R$,
respectively. In particular, it involves the same factors $\zeta_n$.
Due to this reason, in the following we will restrict our attention to $Z_{\rm np,D}$.

\section{Matrix model prediction}
\label{sec-MQM}

\subsection{MQM in the chiral representation} \label{subsec-MQM}

Two-dimensional string theory has a dual description in terms of MQM
in the double scaling limit where it reduces to a system of free fermions in the inverse oscillator potential
with the Fermi energy level identified with the inverse string coupling
\cite{Brezin:1989ss,Ginsparg:1990as,Gross:1990ay}.
The simplest way to obtain the dynamics of the theory is to work 
in the chiral representation where the canonically conjugate variables
are the chiral coordinates in the phase space \cite{Alexandrov:2002fh,Alexandrov:2003ut}
\be
\xpm=\frac{x\pm p}{\sqrt{2}}\, ,
\qquad
\{ \xm,\xp\}=1\, .
\label{lcx}
\ee
The advantage of this chiral representation is that the one-fermion Hamiltonian 
$H_0=\hf(p^2-x^2)$ becomes a {\it first order} linear differential operator
\be
\hat H_0^\pm=-\hf\(\hat x_+\hat x_- +\hat x_-\hat x_+\)=\mp i\( \xpm\frac{\p}{\p\xpm}+\hf\),
\label{chiralH}
\ee
so that its eigenfunctions take a very simple form
\be
\psepm(\xpm)= \frac{e^{\mp\hf \phi_0(E)}}{\sqrt{2\pi }}\,  \xpm^{ \pm i E-{1\over 2}}\, ,
\qquad E\in \IR,
\label{enekk1}
\ee
where $\phi_0(E)$ is a constant phase, whereas in the usual coordinate representation 
they are given by the more complicated parabolic cylinder functions \cite{Moore:1991zv}.

All information about the scattering of free fermions, and hence about the S-matrix of 2d string theory
in the linear dilaton background,
is hidden in the relationship between the two chiral representations.
Namely, since $\psep(\xp)$ and $\psem(\xm)$ represent the same physical state in conjugate representations, 
they must be related by a Fourier transform
\be
\hat S  [\psem](\xp)\equiv
\frac{1}{ \sqrt{2\pi}}\int_0^\infty d\xm \, e^{- i\xp\xm} \psem(\xm)=\psep(\xp).
\label{ennormcond1}
\ee
One should regard $\hat S$ as the S-matrix operator, while $\psepm(\xpm)$ 
as describing incoming and outgoing states \cite{Alexandrov:2002fh}.
Substituting the wave functions \eqref{enekk1} into this relation, one obtains
an integral representation for the scattering phase $\phi_0(E)$:
\be 
e^{-i\phi_0(E)}=\frac{1}{\sqrt{2\pi}}\int_0^\infty \frac{d\xm}{\xm}\, e^{-i\xp\xm} (\xp\xm)^{\hf-iE}.
\label{phase0}
\ee 
It is easy to recognize in \eqref{phase0} the standard integral representation of the Gamma function,
which reproduces the standard result \cite{Moore:1991zv}
\be
e^{-i\phi_0(E)}=\frac{e^{-\frac{\pi i}{4}-\frac{\pi}{2} E}}{\sqrt{2\pi}}\,\Gamma\(\hf-iE\).
\label{scatphase0}
\ee

An important fact is that the scattering phase $\phi_0(E)$ \eqref{scatphase0} is not real.
From \eqref{scatphase0}, one easily finds that
\be
\Im \phi_0(E)=-\frac{1}{2}\, \log\(1+e^{2\pi E}\)=\sum_{n=1}^\infty\frac{(-1)^{n}}{2n}\, e^{2\pi n E}.
\label{Im-scatphase}
\ee
The existence of the imaginary part agrees with the well-known fact that the theory is non-unitary.
Since the occupied energy levels have $E<0$, all terms in \eqref{Im-scatphase} are exponentially suppressed 
and can be seen as a manifestation of the tunneling of fermions to the left side of the potential.
Thus, we can take $i\Im \phi_0$ as the non-perturbative part of the scattering phase, while 
its perturbative part coincides with $\Re \phi_0$.

An important remark is that the equations \eqref{ennormcond1}-\eqref{Im-scatphase}
{\it differ} from similar equations in the previous works on this subject \cite{Alexandrov:2004cg,Alexandrov:2023fvb}.
In particular, the operator $\hat S$ defined by \eqref{ennormcond1} appears to be the inverse of the one usually used,
and the imaginary part of the scattering phase has the opposite sign.
We explain the origin of this discrepancy and why we believe that the definitions given here are the correct ones 
in appendix \ref{ap-signs}.

\subsection{Free energy} \label{subsec-free}

The scattering phase determines not only the S-matrix, but also the free energy of the theory.
At finite temperature $1/(2\pi R)$, it is given by the following integral \cite{Alexandrov:2003qk}
\be
\cF(\mu)
= - R \int_{-\infty}^{\infty}dE \, \frac{\phi_0(E)}{1+e^{2\pi R(\mu+ E)}} \, ,
\label{enepart}
\ee
where $\mu$ is the Fermi energy. From this integral representation, one can also derive 
a simple relation
\be
2\sin \frac{\p_\mu }{2R} \cdot \cF(\mu ) = \phi(-\mu) ,
\label{relFphi}
\ee
which will be useful in what follows.

In this work, we are interested only in the non-perturbative part of the free energy.
It can be obtained by substituting $i\Im \phi_0$ \eqref{Im-scatphase} into the integral \eqref{enepart}
and leads to the following double series of non-perturbative terms 
\cite{Alexandrov:2003nn,Alexandrov:2004cg}\footnote{This formula differs by sign 
	from the non-perturbative free energy given in \cite{Alexandrov:2003nn,Alexandrov:2004cg}.
    This sign difference is due to the sign flip of $\Im \phi_0(E)$ compared to the previous literature,
    explained in the end of the previous subsection and in appendix \ref{ap-signs}.} 
\be
\cF_{\rm np}(\mu)=\sum_{n=1}^\infty {(-1)^{n}\, e^{-2\pi n \mu}\over 4i n\sin{\pi n\over R}}
+\sum_{n=1}^\infty{(-1)^{n} \, e^{-2\pi R n \mu} \over 4i n \sin(\pi R n )}\, .
\label{encFnp}
\ee
The two series have an obvious interpretation as generated by two types of branes in 2d string theory: 
the first series comes from D-instantons having a ZZ-boundary condition for the Liouville field
and Dirichlet on the free boson, while the second series corresponds to D0-branes where 
the Dirichlet condition is replaced by Neumann.
In the following, we will ignore the second contribution. It is absent in the scattering phase since 
it is annihilated by the finite difference operator in \eqref{relFphi}, but in the free energy
it can always be restored by applying T-duality to the D-instanton contribution.
Note, however, that it is crucial to take it into account
if one considers the case of rational $R$ since the apparent singularities of the two terms in \eqref{encFnp}
mutually cancel and the free energy stays finite \cite{Kaushik:2025neu}.

Taking into account that in our normalization $2\pi\mu=1/g_s$ and comparing the exponential in the first series in \eqref{encFnp} 
with the instanton action \eqref{instact}, we see that the $n$-th term corresponds to a contribution 
of a single $(1,n)$ ZZ-brane. However, it can equally arise from $n$ (1,1) ZZ-branes or other combinations of $N_i$ $(1,n_i)$ ZZ-branes
with $\sum_i N_i n_i=n$. This is the above mentioned degeneracy which makes it impossible to distinguish
between different types of instantons. In particular, there is no obvious way to read off the multipliers $\zeta_n$
from the exact non-perturbative free energy.

In fact, if one could calculate the integrals \eqref{defIn}, then it would be possible 
to compare the resulting function \eqref{Znp-res} 
with the exponential of \eqref{encFnp} and to determine all $\zeta_n$ iteratively.
Essentially, this is the approach followed in \cite{Balthazar:2019ypi} where the comparison
with the MQM S-matrix allowed to get the full normalization factors of string D-instanton amplitudes.\footnote{At the time
when that paper was written, it was still unclear how to calculate the annulus amplitudes 
ending on the same ZZ-instanton. Therefore, their result (see \eqref{cN-BRY}) is the full normalization factor 
proportional to the product of $\zeta_n$ and the factor \eqref{one-an} in the $R\to\infty$ limit.}
The problem however is that, as explained below \eqref{Znp-res}, these integrals require a contour prescription 
which is a priory unknown. In \cite{Balthazar:2019ypi}, the so-called Lorentzian prescription has been applied,
but there are other possibilities. For example, in \cite{Sen:2020ruy} another prescription has been advocated
as it was shown to be the only prescription that is consistent with unitarity at the non-perturbative level.
Of course, the bosonic 2d string theory is non-unitary, which makes the situation even more involved.
Given this ambiguity, we prefer {\it not to assume} any particular contour prescription, 
but rather {\it to determine} it form the result for the multipliers $\zeta_n$.

\subsection{Instanton contributions to the scattering phase}
\label{subsec-instcontr}

Note that, due to $\cIi{1}_n=1$, the multiplier $\zeta_1$ can be determined unambiguously
by comparing with the $n=1$ term in \eqref{encFnp} and equals 1/2 \cite{Sen:2020ruy}.
This value is explained by the fact that the integration contour in the path integral turns at 
$\pi/2$ after reaching the saddle responsible for the instanton effect so that 
it follows only a half of the steepest descent contour.
In fact, exactly the same explanation was given to the factor $1/2$ in the $n=1$ term in \eqref{Im-scatphase} \cite{Alexandrov:2004cg}.
This shows that the same factors $\zeta_n$ are expected to arise in the matrix model calculation of non-perturbative effects.

Furthermore, in \cite{Alexandrov:2023fvb} it was noticed that the saddle points of the integral \eqref{phase0}
representing the scattering phase 
are in the one-to-one correspondence with the saddle points of the string theory path integral
corresponding to contributions of $(1,n)$ ZZ-instantons.
This became possible because \cite{Alexandrov:2023fvb} studied 2d string theory in a non-trivial tachyon background 
realized in MQM by changing the asymptotic boundary conditions on the chiral wave functions
so that they represent coherent states of incoming and outgoing free fermions \cite{Alexandrov:2002fh}. 
In such perturbed backgrounds, the degeneracy between one $(1,n)$ and $n$ (1,1) ZZ-instantons is removed \cite{Alexandrov:2003un}
and they turn out to be in the one-to-one correspondence with the so-called double points of the complex curve \cite{Alexandrov:2004ks},
which in turn characterize the saddle points of the integral for the scattering phase \cite{Alexandrov:2004cg,Alexandrov:2023fvb}.
Thus, we can use this integral to disentangle the contributions of different ZZ-branes.

To this end, let us evaluate \eqref{phase0} at $E=-\mu$,
in the large $\mu$ limit at two leading orders.
After the change of variables 
\be  
\xpm=\sqrt{\mu} e^{\pm \tau_\pm},
\ee 
the integral takes the form
\be 
e^{-i\phi_0(-\mu)}=\frac{e^{\(i\mu+\hf\)\log\mu}}{\sqrt{2\pi}}
\int d\tau_-\, e^{-i\mu e^{\tau_+-\tau_-}+\(i\mu+\hf\)(\tau_+-\tau_-)}.
\label{phase01}
\ee 
The saddle point equation extracted from the leading order in $\mu$
\be  
e^{\tau_+-\tau_-}=1
\ee 
has a set of solutions labeled by an integer $n\in\IZ$
\be
\tau_- = \tau_+ - 2\pi i n, 
\label{saddletau}
\ee
which can be identified with the label of $(1,n)$ ZZ-brane.
In \cite[Ap.C]{Harlow:2011ny} it was shown that only the saddle points with $n\ge 0$ contribute 
to the integral.\footnote{More precisely, it was shown that the Gamma function $\Gamma(z)$,
	defined for real and positive $z$ by an integral of type \eqref{phase01} along the real axis, 
	after analytic continuation to the region $\Re z<0$, $\Im z>0$ 
	is given by the same integral along a contour that can be represented as 
	a sum of steepest descent contours of the saddle points \eqref{saddletau} with $n\geq 0$. 
	According to \eqref{scatphase0}, we are interested in analytic continuation to $\Re z\approx 0$, $\Im z>0$
	(recall that $E=-\mu<0$). Since the sign of the imaginary part is the same as in \cite{Harlow:2011ny},
	one gets contributions from the same set of saddle points, 
	while the fact that we are close to the Stokes line 
	$\Re z=0$ is responsible for the coefficients $\zeta_n$ (see section \ref{subsec-median}).}
Therefore, its saddle point approximation yields 
\be
e^{-i\phi_0(-\mu)}= e^{-i\phi_0^{\rm pert}}\sum_{n=0}^\infty \zeta_n (-1)^n e^{-2\pi n\mu} ,
\label{fullphase}
\ee
where
\be  
\phi_0^{\rm pert}\approx \mu(1-\log\mu)+\frac{\pi}{4}\, .
\ee 
In \eqref{fullphase}, we included numerical factors $\zeta_n$ which provide the weight of each saddle point
and are supposed to be the same as the multipliers in \eqref{defN}. 
The perturbative saddle corresponding to $n=0$ has weight $\zeta_0=1$.
Note that the higher order corrections in $1/\mu$ expansion, which we ignored in this calculation,
do not depend on the saddle point and therefore affect only the perturbative contribution $\phi_0^{\rm pert}$
to the scattering phase.
In contrast, the instanton contribution given by the second factor in \eqref{fullphase} is {\it exact},
which agrees with the fact that the relation \eqref{relFphi} and the non-perturbative part of the free energy 
$\cF_{\rm np}$ \eqref{encFnp} are also exact.

The sum over $n$ in \eqref{fullphase} should reproduce the non-perturbative part of the scattering phase
given by its imaginary part \eqref{Im-scatphase}. Equating this sum to $e^{\Im\phi_0}$,
we arrive at a simple condition 
\be 
\sum_{n=0}^\infty \zeta_n x^n =(1-x)^{-1/2},
\label{cond}
\ee
where we denoted $x=-e^{-2\pi \mu}$.
This immediately gives a prediction for the coefficients $\zeta_n$
\be  
\zeta_n=\frac{(2n)!}{2^{2n}(n!)^2}=\frac{(2n-1)!!}{(2n)!!}\, .
\label{reszeta}
\ee

\subsection{Median resummation}
\label{subsec-median}

According to \eqref{reszeta}, we have $\zeta_1=1/2$, $\zeta_2=3/8$, $\zeta_3=5/16$, etc.
While the first number has a clear interpretation in terms of integration contours, 
it is not clear how other rational numbers could arise in a similar way.
On the other hand, we have obtained them by matching the saddle point evaluation of the integral representation of 
the Gamma function \eqref{scatphase0} against the exact expression for its absolute value \eqref{Im-scatphase}.
Thus, they must be an inherent feature of the Gamma function.

The clue is the relation \eqref{cond}. To see how it arises, one should employ the resurgence theory 
(see \cite{resurgbook,sauzin2014introduction1summabilityresurgence,Aniceto:2011nu,Marino:2012zq} for reviews
and \cite{Pasquetti:2010bps,Harlow:2011ny} for a discussion of the Stokes phenomenon for the Gamma function in physics context).
It is relevant for our problem because we are interested in the Gamma function evaluated (in the large $\mu$ limit)
at a pure imaginary argument and the imaginary axis coincides with the Stokes line where the asymptotic expansion
of the Gamma function changes discontinuously. 
The asymptotic expansions on the two sides of the imaginary axis are related by a Stokes factor
\be
{\rm AE}[\Gamma]_-(z)=S(z)\, {\rm AE}[\Gamma]_+(z),
\qquad 
S(z)=\frac{1}{1-e^{\pm 2\pi iz}}\, ,
\label{Stokes}
\ee
where the two signs correspond to the positive and negative imaginary half-axes, respectively.
The reason for this jump is that the steepest descent contour of the perturbative saddle
crosses other saddles situated along the imaginary axis at $z=2\pi i n$, $n\in\IZ$ (cf. \eqref{saddletau}). 
Equivalently, the Borel transform of the asymptotic expansion has singularities along the imaginary axis
so that the inverse Borel transform gives different results, depending on whether the integration contour goes 
from the left or from the right of the singularities, and becomes ambiguous if one insists on
integrating along the Stokes line.

On the other hand, the Gamma function does not have any singularities along the imaginary axis and therefore 
must be well-defined there. 
A resolution of this puzzle is that the non-perturbative ambiguity is canceled between 
various instanton sectors provided along the Stokes line one uses the so-called {\it median resummation}
\cite{Marino:2008ya,Aniceto:2013fka}.\footnote{We thank Marcos Mari\~no for suggesting the relevance of median resummation
	to our problem.} 
For a function $f(z)$ having Borel resummations $\cB_\pm[f]$
related by a Stokes factor $S(z)$ along a Stokes line $\ell$, it amounts to the following prescription
\be 
f(z)=S^{1/2}\cB_+[f]=S^{-1/2} \cB_-[f],
\qquad z\in \ell.
\ee 
In our case, $\cB_+[\Gamma]$ coincides with the contribution of the perturbative saddle and $z=i\mu+\hf$.
Taking into account the form of the Stokes factor \eqref{Stokes},
one concludes that the perturbative contribution should be multiplied by the factor 
$(1+e^{-2\pi\mu})^{-1/2}$. This is precisely the factor in \eqref{cond} generating the multipliers $\zeta_n$.

\subsection{Disentangling instantons}
\label{subsec-disentangle}

Having found the multipliers $\zeta_n$, one can now disentangle contributions of different ZZ-instantons 
to the free energy and the partition function, and thereby produce predictions for string amplitudes.
Applying the relation \eqref{relFphi} to the expression \eqref{fullphase}, one finds that the first term 
in \eqref{encFnp} can be written as
\be
\begin{split} 
\cF_{\rm np,D}(\mu)=&\, \frac{i}{2}\[\sin\frac{\p_\mu}{2R}\]^{-1}\cdot \log\(\sum_{n=0}^\infty \zeta_n (-1)^n e^{-2\pi n\mu}\)
\\
=&\, \frac{i}{2}\sum_{n=1}^\infty (-1)^{n}\, c_n\,  \frac{e^{-2\pi n \mu}}{\sin{\pi n\over R}}\, ,
\end{split} 
\label{encFnpZZ}
\ee
where the coefficients $c_n$ are obtained by expanding the logarithm and rearranging the summations.
As a result, one obtains
\be  
c_n=\sum_{k=1}^n \frac{(-1)^{k}}{k}\sum_{\sum_{j=1}^k n_j=n} \prod_{j=1}^k\zeta_{n_j}\, ,
\label{res-cn}
\ee 
where the sum goes over all ordered integer partitions of $n$.
Substituting \eqref{reszeta}, one can check that $c_n=-\frac{1}{2n}$ consistently with \eqref{encFnp}.
Although this manipulation may look like a mere consistency check, the result \eqref{res-cn}
provides a precise prediction for the contribution of a configuration of $(1,n_j)$ ZZ-instantons, with $j=1,\dots,k$,
to the free energy.

To be able to compare with the string amplitude calculation in \eqref{Znp-res}, we need instead the partition function. 
It is obtained by exponentiating \eqref{encFnpZZ}, which gives
\be  
e^{\cF_{\rm np,D}(\mu)}
= 1+\sum_{n=1}^\infty \cZ_n\,  e^{-2\pi n \mu},
\label{expF}
\ee  
where
\be  
\begin{split} 
\cZ_n=&\, (-1)^n \sum_{m=1}^n\frac{i^{m}}{2^m m!}\sum_{\sum_{l=1}^m n_l=n}  
\prod_{l=1}^m\frac{c_{n_l}}{\sin{\pi n_l\over R}}
\\
=&\,   \sum_{N=1}^n (-1)^{n-N} \sum_{\sum_{i=1}^N n_i=n}  \Ci{N}_\bfn
\prod_{i=1}^N\zeta_{n_i} \, .
\end{split} 
\ee 
The coefficients appearing in the final form are given by 
\be  
\Ci{N}_\bfn=\sum_{m=1}^N\frac{i^m}{2^m m!} \sum_{\sum_{l=1}^m M_l=N} 
\prod_{l=1}^m \frac{1}{M_l\sin{\pi \mathfrak{n}_l\over R}}\, ,
\label{def-Cbfn}
\ee
where $\bfn=(n_1,\dots, n_N)$ as in section \ref{sec-instanton-string} 
and $\mathfrak{n}_l=\sum^{\sum_{s=1}^l M_s}_{i=\sum_{s=1}^{l-1} M_s+1} n_i$.

For ease of comparison with \eqref{Znp-res}, one can also rewrite \eqref{expF} 
as a sum over the number of instantons 
\be 
e^{\cF_{\rm np,D}(\mu)}=1+\sum_{N=1}^\infty \prod_{i=1}^N \[
\sum_{n_i=1}^\infty (-1)^{n_i-1}\zeta_{n_i}\, e^{-2\pi n_i\mu} \]\Ci{N}_\bfn.
\ee 
Collecting the terms with the same product of the multipliers $\zeta_n$ and equating the result to 
the similar contribution in \eqref{Znp-res}, one arrives at a matrix model prediction for 
the integrals of annulus amplitudes between ZZ-instantons
\be 
\cIi{N}_\bfn=(-2i)^N \prod_{i=1}^N \sin\frac{\pi n_i}{R}
\sum_{\sigma\in \cS_N}C_{\sigma(\bfn)},
\label{cIfromMQM-1}
\ee 
where $\cS_N$ is the group of permutations on $N$ elements.
This formula can be further simplified by noticing that the sum over partitions of $N$ in \eqref{def-Cbfn}
produces $m!/|\Sym\{M_l\}|$ identical terms, where $\Sym$ denotes the symmetry group of a finite set and 
$| \cdot |$ denotes the cardinality, 
while the sum over permutations in \eqref{cIfromMQM-1} gives an additional factor of $|\Sym\{M_l\}|\prod_{l=1}^m M_l!$.
Thus, one obtains
\be 
\cIi{N}_\bfn= 
\sum_{m=1}^N (-2i)^{N-m}\sum_{\cup_{l=1}^m \IS_l=\IZ_N \atop \IS_l\cap\IS_k=\emptyset} 
\prod_{l=1}^m \frac{(|\IS_l|-1)!}{\sin\(\frac{\pi}{R}\sum_{i\in\IS_l}n_i\)}
\prod_{i=1}^N \sin\frac{\pi n_i}{R}\, ,
\label{cIfromMQM}
\ee 
where $\IZ_N=\{1,\dots, N\}$ and the sum goes over all disjoint (unordered) partitions into subsets.
A few explicit examples are:
\begin{subequations}
\bea
\cIi{1}_{n} &=& 1 ,
\label{I1}
\\
\cIi{2}_{n_1,n_2} &=& 1-2i\,\frac{ \sin\frac{\pi n_1}{R}\, \sin\frac{\pi n_2}{R}}{\sin\frac{\pi (n_1+n_2)}{R}}\, ,
\label{I2}
\\
\cIi{3}_{n_1,n_2,n_3} &=& 1-2i\(\frac{ \sin\frac{\pi n_1}{R}\, \sin\frac{\pi n_2}{R}}{\sin\frac{\pi (n_1+n_2)}{R}}
+\frac{ \sin\frac{\pi n_1}{R}\, \sin\frac{\pi n_3}{R}}{\sin\frac{\pi (n_1+n_3)}{R}}
+\frac{ \sin\frac{\pi n_2}{R}\, \sin\frac{\pi n_3}{R}}{\sin\frac{\pi (n_2+n_3)}{R}} \)
\nn\\
&&
-8\, \frac{ \sin\frac{\pi n_1}{R}\, \sin\frac{\pi n_2}{R}\, \sin\frac{\pi n_3}{R}}{\sin\frac{\pi (n_1+n_2+n_3)}{R}}\, .
\label{I3}
\eea 
\end{subequations}

\section{Multi-instanton string amplitudes}
\label{sec-string}

\subsection{Contour prescriptions}
\label{subsec-prescrip}

As has been emphasized several times, the integrals \eqref{defIn} determining the multi-instanton string amplitudes
require a contour prescription. The poles of their integrands arise when the stretched open string mode between two
ZZ-instantons becomes on-shell. Before the compactification, this effect is captured by the factor
\be
\frac{1}{(\Delta x)^2-\pi^2(n+m)^2}\, ,
\label{pole}
\ee 
where $\Delta x$ is the Euclidean time distance between the instantons with $(1,n)$ and $(1,m)$ ZZ boundary conditions. 
In \cite{Balthazar:2019ypi}, it was suggested that a natural prescription for the integration contour 
is to take it to be an analytic continuation of the one in Lorentzian theory where $(\Delta x)^2$ 
is negative and the pole is absent. This is equivalent to replacing \eqref{pole} by
\be
\frac{1}{(\Delta x)^2-\pi^2(n+m)^2+i\eps'}\approx \frac{1}{(\Delta x-\pi(n+m) +i\eps)(\Delta x+\pi(n+m) -i\eps)}
\label{pole-eps}
\ee 
with positive infinitesimal parameter $\eps'=2\pi(n+m)\eps$.
The compactification on a circle of radius $R$ amounts to shifting $\Delta x$ by $2\pi R j$ with $j\in\IZ$.
Then taking the product over $j$ replaces each of the factors in \eqref{pole-eps} by a sine function
as in \eqref{def-gmn} with additional $i\eps$ shifts
\be 
\frac{1}{\sin\pi\(y-\frac{n+m}{2R}+i\eps\)\sin\pi\(y+\frac{n+m}{2R}-i\eps\)}\, .
\label{epsgmn}
\ee 
Thus, the Lorentzian prescription of \cite{Balthazar:2019ypi} amounts to avoiding the poles 
at $y=\frac{n+m}{2R}+2\pi k$, $k\in\IZ$, from above and those at  $y=-\frac{n+m}{2R}+2\pi k$ from below.

On the other hand, in \cite{Sen:2020ruy} it was shown that the only prescription compatible with unitarity
is obtained by averaging over the Lorentzian prescription and its complex conjugate where all shifts by $i\eps$
are replaced by $-i\eps$. Due to non-unitarity of the bosonic 2d string theory, the argument in favor of this unitary
prescription is not really applicable, and we have to rely on comparison with MQM to decide which prescription
is the correct one.

We use the prediction \eqref{cIfromMQM} for such comparison.
In fact, already the $N=2$ result \eqref{I2} is sufficient to discriminate between the Lorentzian and unitary prescriptions. 
Indeed, the integrand of $\cIi{2}_{n_1,n_2}$ depends only on one variables $y_{12}$ so that the integral can be easily evaluated.
As a result, one finds \cite{Kaushik:2025neu} that the Lorentzian prescription reproduces \eqref{I2},
whereas the unitary one, equivalent in this case to the principle value prescription, gives the integral equal to 1. 
Thus, it is the Lorentzian prescription that appears to be consistent with the matrix model predictions.

\subsection{Integral evaluation}
\label{subsec-evalint}

Although the above reasoning allows us to decide which of the two prescriptions is favored by MQM, 
it does not prove yet the relevance of the Lorentzian prescription for all multi-instanton amplitudes.
One could imagine that it leads to results different from \eqref{cIfromMQM} for $N>2$ and 
instead there are more complicated prescriptions that are compatible with this prediction.
In this section we show that this is not the case. Remarkably, it turns out that all integrals $\cIi{N}_\bfn$
can be computed exactly and the result of this evaluation in the Lorentzian prescription
precisely coincides with \eqref{cIfromMQM}.

To compute the integrals $\cIi{N}_\bfn$ \eqref{defIn},
let us change the integration variables $y_i\to z_i=e^{2\pi i y_i}$, which maps the integrals to 
\be  
\cIi{N}_\bfn=\oint \prod_{i=1}^N \frac{d z_i}{2\pi i z_i} \,  \prod_{i<j} \gsf_{n_i,n_j}(z_i,z_j),
\label{defInz}
\ee 
where
\be 
\gsf_{n_1,n_2}(z_1,z_2)=
\frac{(\lambda_1 z_1-\lambda_2 z_2)(\lambda_2 z_1-\lambda_1z_2)}{(z_1-\lambda_1\lambda_2 z_2)(\lambda_1\lambda_2 z_1-z_2)}
\label{def-sfg}
\ee 
and $\lambda_i=e^{\pi i n_i/R}$.
The contour in \eqref{defInz} runs over the unit circle in the counter-clockwise direction.
According to the Lorentzian prescription, as follows from \eqref{epsgmn}, 
the factor $\lambda_1\lambda_2$ in the denominator of \eqref{def-sfg}
should be multiplied by $e^{\eps}$, which means that the poles at $z_i=(\lambda_i\lambda_j)z_j$
are avoided from inside the unit circle, whereas the poles at $z_i=(\lambda_i\lambda_j)^{-1}z_j$
are avoided from outside. 

Using this information, one can evaluate the integral \eqref{defInz} by residues.
While the residue at $z_i=0$ or $\infty$ simply removes all $z_i$-dependent factors, 
the contributions of other residues can be evaluated using the relation
\be 
	\Res{z_i=(\lambda_i\lambda_j)^{\pm 1} z_j} \frac{1}{z_i}\, \gsf_{n_i,n_j}(z_i,z_j)\gsf_{n_i,n_k}(z_i,z_k)\gsf_{n_j,n_k}(z_j,z_k)
	= \mp \Lami{2}_{n_i,n_j}\gsf_{n_j+n_i,n_k}(\lambda_i^{\pm 1} z_j,z_k),
\ee 
where we introduced the function 
\be  
\Lami{N}_{\bfn}=\frac{\prod_{i=1}^N(1-\lambda_i^2)}{1-\prod_{i=1}^N\lambda_i^2}
=(-2i)^{N-1}\, \frac{ \prod_{i=1}^N\sin\frac{\pi n_i}{R}}{\sin\(\frac{\pi }{R}\sum_{i=1}^N n_i\)}\, .
\label{def-Lami}
\ee
This function satisfies an important property
\be
\Lami{m}_{\mathfrak{n}}\prod_{l=1}^m \Lami{M_l}_{\bfn_l}=\Lami{N}_{\bfn},
\label{prop-funL}
\ee 
where $N=\sum_{l=1}^m M_l$, 
$\mathfrak{n}=\(\sum^{j_1}_{i=1} n_i,\dots, \sum^{j_m}_{i=j_{m-1}+1} n_i\)$,
$\bfn_l=(n_{j_{l-1}+1},\dots,n_{j_l})$
and $j_l=\sum_{s=1}^l M_s$.
As a result, taking the integral over $z_N$, one finds the following recursion relation
\be  
\cIi{N}_{\bfn}=\cIi{N-1}_{\bfn}+
\sum_{i=1}^{N-1} \Lami{2}_{n_i,n_N}\,\cIi{N-1}_{\bfn[i]},
\label{defInz-L}
\ee 
where we denoted $\bfn[i]=(n_1,\dots,n_i+n_N,\dots n_{N-1})$.
On the r.h.s. the first term comes from the residue at $z_N=0$ (or $z_N=\infty$), while the second term is generated 
by residues at $z_N=(\lambda_i\lambda_j)^{-1}z_i$ (or $z_N=\lambda_i\lambda_j z_i$).

Using \eqref{prop-funL} and \eqref{defInz-L}, it is now easy to prove by induction that $\cIi{N}_{\bfn}$
in the Lorentzian prescription agrees with the prediction \eqref{cIfromMQM}, which can be equivalently written as
\be 
\cIi{N}_{\bfn}= 
\sum_{m=1}^N \sum_{\cup_{l=1}^m \IS_l=\IZ_N \atop \IS_l\cap\IS_k=\emptyset} 
\prod_{l=1}^m (|\IS_l|-1)!\, \Lami{|\IS_l|}_{\mathfrak{n}_l}\, ,
\label{cIfromMQM-set}
\ee
where $\mathfrak{n}_l=\{n_i\}_{i\in\IS_l}$.
Indeed, let us assume that this formula holds for $N-1$ and prove it for $N$.
Applying the recursion relation \eqref{defInz-L} and the induction hypothesis, one obtains 
\be 
\cIi{N}_{\bfn}= 
\sum_{m=1}^{N-1} \sum_{\cup_{l=1}^m \IS_l=\IZ_{N-1} \atop \IS_l\cap\IS_k=\emptyset} 
\prod_{l=1}^m (|\IS_l|-1)!\, \Lami{|\IS_l|}_{\mathfrak{n}_l}
+\sum_{i=1}^{N-1} \Lami{2}_{n_i,n_N}\,
\sum_{m=1}^{N-1} \sum_{\cup_{l=1}^m \IS_l=\IZ_{N-1} \atop \IS_l\cap\IS_k=\emptyset} 
\prod_{l=1}^m (|\IS_l|-1)!\, \Lami{|\IS_l|}_{\mathfrak{n}_l[i]}\, ,
\label{cIfromMQM-proof}
\ee
where $\mathfrak{n}_l[i]$ denotes the subset where, if $i\in \IS_l$, $n_i$ is replaced by $n_i+n_N$.
The first term can be immediately identified as the contribution to \eqref{cIfromMQM-set} 
corresponding to partitions where one of the subsets $\IS_l$ coincides with $\{N\}$.
To see that the second term captures the rest, let us consider a term in \eqref{cIfromMQM-set} where $\{N\}\in\IS_{l_0}$
and $\IS'_{l_0}=\IS_{l_0}\backslash \{N\}$ is non-empty. We claim that it is equal to the sum of contributions to the second term 
in \eqref{cIfromMQM-proof} with $i\in\IS'_{l_0}$ and partition $\IZ_{N-1}=\(\cup_{l\ne l_0} \IS_l\)\cup\IS'_{l_0}$.
Indeed, the property \eqref{prop-funL} ensures that 
$\Lami{2}_{n_i,n_N}\Lami{|\IS'_{l_0}|}_{\mathfrak{n}_{l_0}[i]}=\Lami{|\IS_{l_0}|}_{\mathfrak{n}_{l_0}}$.
As a result, all terms in the sum become identical. This produces the factor $|\IS_{l_0}|-1$, 
which upgrades the factor $(|\IS'_{l_0}|-1)!$ to $(|\IS_{l_0}|-1)!$, as required.
This proves the representation \eqref{cIfromMQM-set} and hence verifies the MQM prediction.

\subsection{Free energy and SFT effective action}

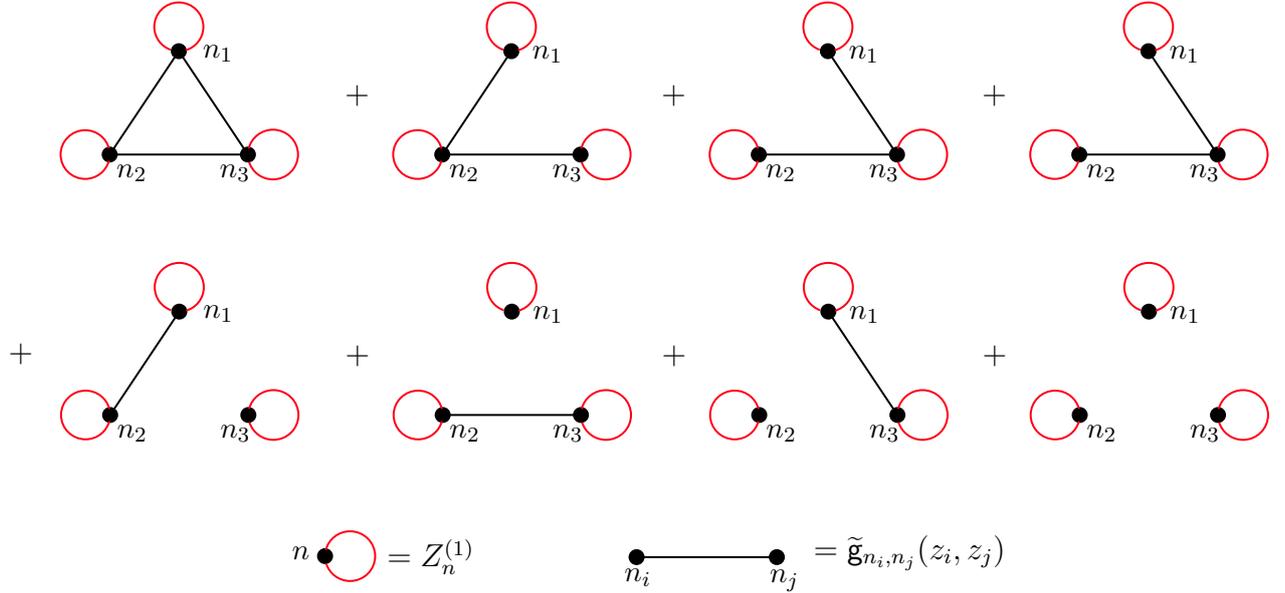
\begin{figure}[t]
	\centering
	\tikzset{every picture/.style={line width=0.75pt}} %set default line width to 0.75pt        	
	\begin{tikzpicture}[x=0.75pt,y=0.75pt,yscale=-1,xscale=1]
		%uncomment if require: \path (0,333); %set diagram left start at 0, and has height of 333
		
		%Shape: Circle [id:dp6896537085758293] 
		\draw  [fill={rgb, 255:red, 0; green, 0; blue, 0 }  ,fill opacity=1 ] (294.5,94.5) .. controls (294.5,92.57) and (296.07,91) .. (298,91) .. controls (299.93,91) and (301.5,92.57) .. (301.5,94.5) .. controls (301.5,96.43) and (299.93,98) .. (298,98) .. controls (296.07,98) and (294.5,96.43) .. (294.5,94.5) -- cycle ;
		%Shape: Circle [id:dp6229337543802156] 
		\draw  [fill={rgb, 255:red, 0; green, 0; blue, 0 }  ,fill opacity=1 ] (260,42.5) .. controls (260,40.57) and (261.57,39) .. (263.5,39) .. controls (265.43,39) and (267,40.57) .. (267,42.5) .. controls (267,44.43) and (265.43,46) .. (263.5,46) .. controls (261.57,46) and (260,44.43) .. (260,42.5) -- cycle ;
		%Shape: Circle [id:dp8090042923272662] 
		\draw  [fill={rgb, 255:red, 0; green, 0; blue, 0 }  ,fill opacity=1 ] (225.5,94.5) .. controls (225.5,92.57) and (227.07,91) .. (229,91) .. controls (230.93,91) and (232.5,92.57) .. (232.5,94.5) .. controls (232.5,96.43) and (230.93,98) .. (229,98) .. controls (227.07,98) and (225.5,96.43) .. (225.5,94.5) -- cycle ;
		%Shape: Arc [id:dp8343456808536913] 
		\draw  [draw opacity=0] (228.6,97.61) .. controls (227.23,102.87) and (222.44,106.75) .. (216.75,106.75) .. controls (209.98,106.75) and (204.5,101.27) .. (204.5,94.5) .. controls (204.5,87.73) and (209.98,82.25) .. (216.75,82.25) .. controls (222.47,82.25) and (227.27,86.17) .. (228.62,91.47) -- (216.75,94.5) -- cycle ; \draw  [color={rgb, 255:red, 255; green, 0; blue, 24 }  ,draw opacity=1 ] (228.6,97.61) .. controls (227.23,102.87) and (222.44,106.75) .. (216.75,106.75) .. controls (209.98,106.75) and (204.5,101.27) .. (204.5,94.5) .. controls (204.5,87.73) and (209.98,82.25) .. (216.75,82.25) .. controls (222.47,82.25) and (227.27,86.17) .. (228.62,91.47) ;  
		%Shape: Arc [id:dp8055747576461817] 
		\draw  [draw opacity=0] (259.39,41.79) .. controls (254.65,40.1) and (251.25,35.57) .. (251.25,30.25) .. controls (251.25,23.48) and (256.73,18) .. (263.5,18) .. controls (270.27,18) and (275.75,23.48) .. (275.75,30.25) .. controls (275.75,35.85) and (272,40.57) .. (266.87,42.03) -- (263.5,30.25) -- cycle ; \draw  [color={rgb, 255:red, 255; green, 0; blue, 24 }  ,draw opacity=1 ] (259.39,41.79) .. controls (254.65,40.1) and (251.25,35.57) .. (251.25,30.25) .. controls (251.25,23.48) and (256.73,18) .. (263.5,18) .. controls (270.27,18) and (275.75,23.48) .. (275.75,30.25) .. controls (275.75,35.85) and (272,40.57) .. (266.87,42.03) ;  
		%Shape: Arc [id:dp7838053493959638] 
		\draw  [draw opacity=0] (298.37,91.47) .. controls (299.73,86.03) and (304.64,82) .. (310.5,82) .. controls (317.4,82) and (323,87.6) .. (323,94.5) .. controls (323,101.4) and (317.4,107) .. (310.5,107) .. controls (304.87,107) and (300.11,103.28) .. (298.55,98.16) -- (310.5,94.5) -- cycle ; \draw  [color={rgb, 255:red, 255; green, 0; blue, 24 }  ,draw opacity=1 ] (298.37,91.47) .. controls (299.73,86.03) and (304.64,82) .. (310.5,82) .. controls (317.4,82) and (323,87.6) .. (323,94.5) .. controls (323,101.4) and (317.4,107) .. (310.5,107) .. controls (304.87,107) and (300.11,103.28) .. (298.55,98.16) ;  
		%Straight Lines [id:da973902949846259] 
		\draw    (263.5,42.5) -- (229,94.5) ;
		%Straight Lines [id:da42362925085545167] 
		\draw    (229,94.5) -- (298,94.5) ;
		%Shape: Circle [id:dp2173138527105104] 
		\draw  [fill={rgb, 255:red, 0; green, 0; blue, 0 }  ,fill opacity=1 ] (167,296.5) .. controls (167,294.57) and (168.57,293) .. (170.5,293) .. controls (172.43,293) and (174,294.57) .. (174,296.5) .. controls (174,298.43) and (172.43,300) .. (170.5,300) .. controls (168.57,300) and (167,298.43) .. (167,296.5) -- cycle ;
		%Shape: Arc [id:dp565540079679087] 
		\draw  [draw opacity=0] (170.87,293.47) .. controls (172.23,288.03) and (177.14,284) .. (183,284) .. controls (189.9,284) and (195.5,289.6) .. (195.5,296.5) .. controls (195.5,303.4) and (189.9,309) .. (183,309) .. controls (177.37,309) and (172.61,305.28) .. (171.05,300.16) -- (183,296.5) -- cycle ; \draw  [color={rgb, 255:red, 255; green, 0; blue, 24 }  ,draw opacity=1 ] (170.87,293.47) .. controls (172.23,288.03) and (177.14,284) .. (183,284) .. controls (189.9,284) and (195.5,289.6) .. (195.5,296.5) .. controls (195.5,303.4) and (189.9,309) .. (183,309) .. controls (177.37,309) and (172.61,305.28) .. (171.05,300.16) ;  
		%Straight Lines [id:da4181886824990749] 
		\draw    (326,297) -- (395,297) ;
		%Shape: Circle [id:dp6008435125694468] 
		\draw  [fill={rgb, 255:red, 0; green, 0; blue, 0 }  ,fill opacity=1 ] (322.5,297) .. controls (322.5,295.07) and (324.07,293.5) .. (326,293.5) .. controls (327.93,293.5) and (329.5,295.07) .. (329.5,297) .. controls (329.5,298.93) and (327.93,300.5) .. (326,300.5) .. controls (324.07,300.5) and (322.5,298.93) .. (322.5,297) -- cycle ;
		%Shape: Circle [id:dp1439591937664113] 
		\draw  [fill={rgb, 255:red, 0; green, 0; blue, 0 }  ,fill opacity=1 ] (392.5,297) .. controls (392.5,295.07) and (394.07,293.5) .. (396,293.5) .. controls (397.93,293.5) and (399.5,295.07) .. (399.5,297) .. controls (399.5,298.93) and (397.93,300.5) .. (396,300.5) .. controls (394.07,300.5) and (392.5,298.93) .. (392.5,297) -- cycle ;
		%Shape: Circle [id:dp08257992669095937] 
		\draw  [fill={rgb, 255:red, 0; green, 0; blue, 0 }  ,fill opacity=1 ] (128.5,94.5) .. controls (128.5,92.57) and (130.07,91) .. (132,91) .. controls (133.93,91) and (135.5,92.57) .. (135.5,94.5) .. controls (135.5,96.43) and (133.93,98) .. (132,98) .. controls (130.07,98) and (128.5,96.43) .. (128.5,94.5) -- cycle ;
		%Shape: Circle [id:dp44380049214522976] 
		\draw  [fill={rgb, 255:red, 0; green, 0; blue, 0 }  ,fill opacity=1 ] (94,42.5) .. controls (94,40.57) and (95.57,39) .. (97.5,39) .. controls (99.43,39) and (101,40.57) .. (101,42.5) .. controls (101,44.43) and (99.43,46) .. (97.5,46) .. controls (95.57,46) and (94,44.43) .. (94,42.5) -- cycle ;
		%Shape: Circle [id:dp37586403103989396] 
		\draw  [fill={rgb, 255:red, 0; green, 0; blue, 0 }  ,fill opacity=1 ] (59.5,94.5) .. controls (59.5,92.57) and (61.07,91) .. (63,91) .. controls (64.93,91) and (66.5,92.57) .. (66.5,94.5) .. controls (66.5,96.43) and (64.93,98) .. (63,98) .. controls (61.07,98) and (59.5,96.43) .. (59.5,94.5) -- cycle ;
		%Shape: Arc [id:dp4208939644085241] 
		\draw  [draw opacity=0] (62.6,97.61) .. controls (61.23,102.87) and (56.44,106.75) .. (50.75,106.75) .. controls (43.98,106.75) and (38.5,101.27) .. (38.5,94.5) .. controls (38.5,87.73) and (43.98,82.25) .. (50.75,82.25) .. controls (56.47,82.25) and (61.27,86.17) .. (62.62,91.47) -- (50.75,94.5) -- cycle ; \draw  [color={rgb, 255:red, 255; green, 0; blue, 24 }  ,draw opacity=1 ] (62.6,97.61) .. controls (61.23,102.87) and (56.44,106.75) .. (50.75,106.75) .. controls (43.98,106.75) and (38.5,101.27) .. (38.5,94.5) .. controls (38.5,87.73) and (43.98,82.25) .. (50.75,82.25) .. controls (56.47,82.25) and (61.27,86.17) .. (62.62,91.47) ;  
		%Shape: Arc [id:dp7571843612563831] 
		\draw  [draw opacity=0] (93.39,41.79) .. controls (88.65,40.1) and (85.25,35.57) .. (85.25,30.25) .. controls (85.25,23.48) and (90.73,18) .. (97.5,18) .. controls (104.27,18) and (109.75,23.48) .. (109.75,30.25) .. controls (109.75,35.85) and (106,40.57) .. (100.87,42.03) -- (97.5,30.25) -- cycle ; \draw  [color={rgb, 255:red, 255; green, 0; blue, 24 }  ,draw opacity=1 ] (93.39,41.79) .. controls (88.65,40.1) and (85.25,35.57) .. (85.25,30.25) .. controls (85.25,23.48) and (90.73,18) .. (97.5,18) .. controls (104.27,18) and (109.75,23.48) .. (109.75,30.25) .. controls (109.75,35.85) and (106,40.57) .. (100.87,42.03) ;  
		%Shape: Arc [id:dp33710145770865274] 
		\draw  [draw opacity=0] (132.37,91.47) .. controls (133.73,86.03) and (138.64,82) .. (144.5,82) .. controls (151.4,82) and (157,87.6) .. (157,94.5) .. controls (157,101.4) and (151.4,107) .. (144.5,107) .. controls (138.87,107) and (134.11,103.28) .. (132.55,98.16) -- (144.5,94.5) -- cycle ; \draw  [color={rgb, 255:red, 255; green, 0; blue, 24 }  ,draw opacity=1 ] (132.37,91.47) .. controls (133.73,86.03) and (138.64,82) .. (144.5,82) .. controls (151.4,82) and (157,87.6) .. (157,94.5) .. controls (157,101.4) and (151.4,107) .. (144.5,107) .. controls (138.87,107) and (134.11,103.28) .. (132.55,98.16) ;  
		%Straight Lines [id:da15098639635438948] 
		\draw    (97.5,42.5) -- (63,94.5) ;
		%Straight Lines [id:da8307770063524463] 
		\draw    (63,94.5) -- (132,94.5) ;
		%Straight Lines [id:da36513334028075717] 
		\draw    (97.5,42.5) -- (132,94.5) ;
		%Shape: Circle [id:dp3645643695836883] 
		\draw  [fill={rgb, 255:red, 0; green, 0; blue, 0 }  ,fill opacity=1 ] (452.5,94.5) .. controls (452.5,92.57) and (454.07,91) .. (456,91) .. controls (457.93,91) and (459.5,92.57) .. (459.5,94.5) .. controls (459.5,96.43) and (457.93,98) .. (456,98) .. controls (454.07,98) and (452.5,96.43) .. (452.5,94.5) -- cycle ;
		%Shape: Circle [id:dp15087094234723253] 
		\draw  [fill={rgb, 255:red, 0; green, 0; blue, 0 }  ,fill opacity=1 ] (418,42.5) .. controls (418,40.57) and (419.57,39) .. (421.5,39) .. controls (423.43,39) and (425,40.57) .. (425,42.5) .. controls (425,44.43) and (423.43,46) .. (421.5,46) .. controls (419.57,46) and (418,44.43) .. (418,42.5) -- cycle ;
		%Shape: Circle [id:dp6060025547224894] 
		\draw  [fill={rgb, 255:red, 0; green, 0; blue, 0 }  ,fill opacity=1 ] (383.5,94.5) .. controls (383.5,92.57) and (385.07,91) .. (387,91) .. controls (388.93,91) and (390.5,92.57) .. (390.5,94.5) .. controls (390.5,96.43) and (388.93,98) .. (387,98) .. controls (385.07,98) and (383.5,96.43) .. (383.5,94.5) -- cycle ;
		%Shape: Arc [id:dp7949997737173767] 
		\draw  [draw opacity=0] (386.6,97.61) .. controls (385.23,102.87) and (380.44,106.75) .. (374.75,106.75) .. controls (367.98,106.75) and (362.5,101.27) .. (362.5,94.5) .. controls (362.5,87.73) and (367.98,82.25) .. (374.75,82.25) .. controls (380.47,82.25) and (385.27,86.17) .. (386.62,91.47) -- (374.75,94.5) -- cycle ; \draw  [color={rgb, 255:red, 255; green, 0; blue, 24 }  ,draw opacity=1 ] (386.6,97.61) .. controls (385.23,102.87) and (380.44,106.75) .. (374.75,106.75) .. controls (367.98,106.75) and (362.5,101.27) .. (362.5,94.5) .. controls (362.5,87.73) and (367.98,82.25) .. (374.75,82.25) .. controls (380.47,82.25) and (385.27,86.17) .. (386.62,91.47) ;  
		%Shape: Arc [id:dp6510519935047698] 
		\draw  [draw opacity=0] (417.39,41.79) .. controls (412.65,40.1) and (409.25,35.57) .. (409.25,30.25) .. controls (409.25,23.48) and (414.73,18) .. (421.5,18) .. controls (428.27,18) and (433.75,23.48) .. (433.75,30.25) .. controls (433.75,35.85) and (430,40.57) .. (424.87,42.03) -- (421.5,30.25) -- cycle ; \draw  [color={rgb, 255:red, 255; green, 0; blue, 24 }  ,draw opacity=1 ] (417.39,41.79) .. controls (412.65,40.1) and (409.25,35.57) .. (409.25,30.25) .. controls (409.25,23.48) and (414.73,18) .. (421.5,18) .. controls (428.27,18) and (433.75,23.48) .. (433.75,30.25) .. controls (433.75,35.85) and (430,40.57) .. (424.87,42.03) ;  
		%Shape: Arc [id:dp5951598833528232] 
		\draw  [draw opacity=0] (456.37,91.47) .. controls (457.73,86.03) and (462.64,82) .. (468.5,82) .. controls (475.4,82) and (481,87.6) .. (481,94.5) .. controls (481,101.4) and (475.4,107) .. (468.5,107) .. controls (462.87,107) and (458.11,103.28) .. (456.55,98.16) -- (468.5,94.5) -- cycle ; \draw  [color={rgb, 255:red, 255; green, 0; blue, 24 }  ,draw opacity=1 ] (456.37,91.47) .. controls (457.73,86.03) and (462.64,82) .. (468.5,82) .. controls (475.4,82) and (481,87.6) .. (481,94.5) .. controls (481,101.4) and (475.4,107) .. (468.5,107) .. controls (462.87,107) and (458.11,103.28) .. (456.55,98.16) ;  
		%Straight Lines [id:da8442911984536364] 
		\draw    (421.5,42.5) -- (456,94.5) ;
		%Straight Lines [id:da5664760849824798] 
		\draw    (387,94.5) -- (456,94.5) ;
		%Shape: Circle [id:dp7279372687402459] 
		\draw  [fill={rgb, 255:red, 0; green, 0; blue, 0 }  ,fill opacity=1 ] (612.5,94.5) .. controls (612.5,92.57) and (614.07,91) .. (616,91) .. controls (617.93,91) and (619.5,92.57) .. (619.5,94.5) .. controls (619.5,96.43) and (617.93,98) .. (616,98) .. controls (614.07,98) and (612.5,96.43) .. (612.5,94.5) -- cycle ;
		%Shape: Circle [id:dp21949971093145215] 
		\draw  [fill={rgb, 255:red, 0; green, 0; blue, 0 }  ,fill opacity=1 ] (578,42.5) .. controls (578,40.57) and (579.57,39) .. (581.5,39) .. controls (583.43,39) and (585,40.57) .. (585,42.5) .. controls (585,44.43) and (583.43,46) .. (581.5,46) .. controls (579.57,46) and (578,44.43) .. (578,42.5) -- cycle ;
		%Shape: Circle [id:dp8191050681555605] 
		\draw  [fill={rgb, 255:red, 0; green, 0; blue, 0 }  ,fill opacity=1 ] (543.5,94.5) .. controls (543.5,92.57) and (545.07,91) .. (547,91) .. controls (548.93,91) and (550.5,92.57) .. (550.5,94.5) .. controls (550.5,96.43) and (548.93,98) .. (547,98) .. controls (545.07,98) and (543.5,96.43) .. (543.5,94.5) -- cycle ;
		%Shape: Arc [id:dp5686481558139226] 
		\draw  [draw opacity=0] (546.6,97.61) .. controls (545.23,102.87) and (540.44,106.75) .. (534.75,106.75) .. controls (527.98,106.75) and (522.5,101.27) .. (522.5,94.5) .. controls (522.5,87.73) and (527.98,82.25) .. (534.75,82.25) .. controls (540.47,82.25) and (545.27,86.17) .. (546.62,91.47) -- (534.75,94.5) -- cycle ; \draw  [color={rgb, 255:red, 255; green, 0; blue, 24 }  ,draw opacity=1 ] (546.6,97.61) .. controls (545.23,102.87) and (540.44,106.75) .. (534.75,106.75) .. controls (527.98,106.75) and (522.5,101.27) .. (522.5,94.5) .. controls (522.5,87.73) and (527.98,82.25) .. (534.75,82.25) .. controls (540.47,82.25) and (545.27,86.17) .. (546.62,91.47) ;  
		%Shape: Arc [id:dp10600797599412337] 
		\draw  [draw opacity=0] (577.39,41.79) .. controls (572.65,40.1) and (569.25,35.57) .. (569.25,30.25) .. controls (569.25,23.48) and (574.73,18) .. (581.5,18) .. controls (588.27,18) and (593.75,23.48) .. (593.75,30.25) .. controls (593.75,35.85) and (590,40.57) .. (584.87,42.03) -- (581.5,30.25) -- cycle ; \draw  [color={rgb, 255:red, 255; green, 0; blue, 24 }  ,draw opacity=1 ] (577.39,41.79) .. controls (572.65,40.1) and (569.25,35.57) .. (569.25,30.25) .. controls (569.25,23.48) and (574.73,18) .. (581.5,18) .. controls (588.27,18) and (593.75,23.48) .. (593.75,30.25) .. controls (593.75,35.85) and (590,40.57) .. (584.87,42.03) ;  
		%Shape: Arc [id:dp3268997488163212] 
		\draw  [draw opacity=0] (616.37,91.47) .. controls (617.73,86.03) and (622.64,82) .. (628.5,82) .. controls (635.4,82) and (641,87.6) .. (641,94.5) .. controls (641,101.4) and (635.4,107) .. (628.5,107) .. controls (622.87,107) and (618.11,103.28) .. (616.55,98.16) -- (628.5,94.5) -- cycle ; \draw  [color={rgb, 255:red, 255; green, 0; blue, 24 }  ,draw opacity=1 ] (616.37,91.47) .. controls (617.73,86.03) and (622.64,82) .. (628.5,82) .. controls (635.4,82) and (641,87.6) .. (641,94.5) .. controls (641,101.4) and (635.4,107) .. (628.5,107) .. controls (622.87,107) and (618.11,103.28) .. (616.55,98.16) ;  
		%Straight Lines [id:da1922594875707918] 
		\draw    (581.5,42.5) -- (616,94.5) ;
		%Straight Lines [id:da4244147605461003] 
		\draw    (547,94.5) -- (616,94.5) ;
		%Shape: Circle [id:dp4282681108605433] 
		\draw  [fill={rgb, 255:red, 0; green, 0; blue, 0 }  ,fill opacity=1 ] (294.72,225.5) .. controls (294.72,223.57) and (296.29,222) .. (298.22,222) .. controls (300.16,222) and (301.72,223.57) .. (301.72,225.5) .. controls (301.72,227.43) and (300.16,229) .. (298.22,229) .. controls (296.29,229) and (294.72,227.43) .. (294.72,225.5) -- cycle ;
		%Shape: Circle [id:dp9504846105314811] 
		\draw  [fill={rgb, 255:red, 0; green, 0; blue, 0 }  ,fill opacity=1 ] (260.22,173.5) .. controls (260.22,171.57) and (261.79,170) .. (263.72,170) .. controls (265.66,170) and (267.22,171.57) .. (267.22,173.5) .. controls (267.22,175.43) and (265.66,177) .. (263.72,177) .. controls (261.79,177) and (260.22,175.43) .. (260.22,173.5) -- cycle ;
		%Shape: Circle [id:dp9164755886566003] 
		\draw  [fill={rgb, 255:red, 0; green, 0; blue, 0 }  ,fill opacity=1 ] (225.72,225.5) .. controls (225.72,223.57) and (227.29,222) .. (229.22,222) .. controls (231.16,222) and (232.72,223.57) .. (232.72,225.5) .. controls (232.72,227.43) and (231.16,229) .. (229.22,229) .. controls (227.29,229) and (225.72,227.43) .. (225.72,225.5) -- cycle ;
		%Shape: Arc [id:dp5068335050946942] 
		\draw  [draw opacity=0] (228.82,228.61) .. controls (227.45,233.87) and (222.66,237.75) .. (216.97,237.75) .. controls (210.21,237.75) and (204.72,232.27) .. (204.72,225.5) .. controls (204.72,218.73) and (210.21,213.25) .. (216.97,213.25) .. controls (222.69,213.25) and (227.5,217.17) .. (228.84,222.47) -- (216.97,225.5) -- cycle ; \draw  [color={rgb, 255:red, 255; green, 0; blue, 24 }  ,draw opacity=1 ] (228.82,228.61) .. controls (227.45,233.87) and (222.66,237.75) .. (216.97,237.75) .. controls (210.21,237.75) and (204.72,232.27) .. (204.72,225.5) .. controls (204.72,218.73) and (210.21,213.25) .. (216.97,213.25) .. controls (222.69,213.25) and (227.5,217.17) .. (228.84,222.47) ;  
		%Shape: Arc [id:dp13993123590500267] 
		\draw  [draw opacity=0] (259.61,172.79) .. controls (254.87,171.1) and (251.47,166.57) .. (251.47,161.25) .. controls (251.47,154.48) and (256.96,149) .. (263.72,149) .. controls (270.49,149) and (275.97,154.48) .. (275.97,161.25) .. controls (275.97,166.85) and (272.22,171.57) .. (267.09,173.03) -- (263.72,161.25) -- cycle ; \draw  [color={rgb, 255:red, 255; green, 0; blue, 24 }  ,draw opacity=1 ] (259.61,172.79) .. controls (254.87,171.1) and (251.47,166.57) .. (251.47,161.25) .. controls (251.47,154.48) and (256.96,149) .. (263.72,149) .. controls (270.49,149) and (275.97,154.48) .. (275.97,161.25) .. controls (275.97,166.85) and (272.22,171.57) .. (267.09,173.03) ;  
		%Shape: Arc [id:dp7294499989660798] 
		\draw  [draw opacity=0] (298.59,222.47) .. controls (299.95,217.03) and (304.86,213) .. (310.72,213) .. controls (317.63,213) and (323.22,218.6) .. (323.22,225.5) .. controls (323.22,232.4) and (317.63,238) .. (310.72,238) .. controls (305.09,238) and (300.33,234.28) .. (298.77,229.16) -- (310.72,225.5) -- cycle ; \draw  [color={rgb, 255:red, 255; green, 0; blue, 24 }  ,draw opacity=1 ] (298.59,222.47) .. controls (299.95,217.03) and (304.86,213) .. (310.72,213) .. controls (317.63,213) and (323.22,218.6) .. (323.22,225.5) .. controls (323.22,232.4) and (317.63,238) .. (310.72,238) .. controls (305.09,238) and (300.33,234.28) .. (298.77,229.16) ;  
		%Straight Lines [id:da6539228345820312] 
		\draw    (229.22,225.5) -- (298.22,225.5) ;
		%Shape: Circle [id:dp4723212667919219] 
		\draw  [fill={rgb, 255:red, 0; green, 0; blue, 0 }  ,fill opacity=1 ] (128.72,225.5) .. controls (128.72,223.57) and (130.29,222) .. (132.22,222) .. controls (134.16,222) and (135.72,223.57) .. (135.72,225.5) .. controls (135.72,227.43) and (134.16,229) .. (132.22,229) .. controls (130.29,229) and (128.72,227.43) .. (128.72,225.5) -- cycle ;
		%Shape: Circle [id:dp8021339461680244] 
		\draw  [fill={rgb, 255:red, 0; green, 0; blue, 0 }  ,fill opacity=1 ] (94.22,173.5) .. controls (94.22,171.57) and (95.79,170) .. (97.72,170) .. controls (99.66,170) and (101.22,171.57) .. (101.22,173.5) .. controls (101.22,175.43) and (99.66,177) .. (97.72,177) .. controls (95.79,177) and (94.22,175.43) .. (94.22,173.5) -- cycle ;
		%Shape: Circle [id:dp44317600352415054] 
		\draw  [fill={rgb, 255:red, 0; green, 0; blue, 0 }  ,fill opacity=1 ] (59.72,225.5) .. controls (59.72,223.57) and (61.29,222) .. (63.22,222) .. controls (65.16,222) and (66.72,223.57) .. (66.72,225.5) .. controls (66.72,227.43) and (65.16,229) .. (63.22,229) .. controls (61.29,229) and (59.72,227.43) .. (59.72,225.5) -- cycle ;
		%Shape: Arc [id:dp385147436845772] 
		\draw  [draw opacity=0] (62.82,228.61) .. controls (61.45,233.87) and (56.66,237.75) .. (50.97,237.75) .. controls (44.21,237.75) and (38.72,232.27) .. (38.72,225.5) .. controls (38.72,218.73) and (44.21,213.25) .. (50.97,213.25) .. controls (56.69,213.25) and (61.5,217.17) .. (62.84,222.47) -- (50.97,225.5) -- cycle ; \draw  [color={rgb, 255:red, 255; green, 0; blue, 24 }  ,draw opacity=1 ] (62.82,228.61) .. controls (61.45,233.87) and (56.66,237.75) .. (50.97,237.75) .. controls (44.21,237.75) and (38.72,232.27) .. (38.72,225.5) .. controls (38.72,218.73) and (44.21,213.25) .. (50.97,213.25) .. controls (56.69,213.25) and (61.5,217.17) .. (62.84,222.47) ;  
		%Shape: Arc [id:dp8499438513519827] 
		\draw  [draw opacity=0] (93.61,172.79) .. controls (88.87,171.1) and (85.47,166.57) .. (85.47,161.25) .. controls (85.47,154.48) and (90.96,149) .. (97.72,149) .. controls (104.49,149) and (109.97,154.48) .. (109.97,161.25) .. controls (109.97,166.85) and (106.22,171.57) .. (101.09,173.03) -- (97.72,161.25) -- cycle ; \draw  [color={rgb, 255:red, 255; green, 0; blue, 24 }  ,draw opacity=1 ] (93.61,172.79) .. controls (88.87,171.1) and (85.47,166.57) .. (85.47,161.25) .. controls (85.47,154.48) and (90.96,149) .. (97.72,149) .. controls (104.49,149) and (109.97,154.48) .. (109.97,161.25) .. controls (109.97,166.85) and (106.22,171.57) .. (101.09,173.03) ;  
		%Shape: Arc [id:dp995513114069134] 
		\draw  [draw opacity=0] (132.59,222.47) .. controls (133.95,217.03) and (138.86,213) .. (144.72,213) .. controls (151.63,213) and (157.22,218.6) .. (157.22,225.5) .. controls (157.22,232.4) and (151.63,238) .. (144.72,238) .. controls (139.09,238) and (134.33,234.28) .. (132.77,229.16) -- (144.72,225.5) -- cycle ; \draw  [color={rgb, 255:red, 255; green, 0; blue, 24 }  ,draw opacity=1 ] (132.59,222.47) .. controls (133.95,217.03) and (138.86,213) .. (144.72,213) .. controls (151.63,213) and (157.22,218.6) .. (157.22,225.5) .. controls (157.22,232.4) and (151.63,238) .. (144.72,238) .. controls (139.09,238) and (134.33,234.28) .. (132.77,229.16) ;  
		%Straight Lines [id:da24830676282910746] 
		\draw    (97.72,173.5) -- (63.22,225.5) ;
		%Shape: Circle [id:dp5232548936779625] 
		\draw  [fill={rgb, 255:red, 0; green, 0; blue, 0 }  ,fill opacity=1 ] (452.72,225.5) .. controls (452.72,223.57) and (454.29,222) .. (456.22,222) .. controls (458.16,222) and (459.72,223.57) .. (459.72,225.5) .. controls (459.72,227.43) and (458.16,229) .. (456.22,229) .. controls (454.29,229) and (452.72,227.43) .. (452.72,225.5) -- cycle ;
		%Shape: Circle [id:dp7671537718603619] 
		\draw  [fill={rgb, 255:red, 0; green, 0; blue, 0 }  ,fill opacity=1 ] (418.22,173.5) .. controls (418.22,171.57) and (419.79,170) .. (421.72,170) .. controls (423.66,170) and (425.22,171.57) .. (425.22,173.5) .. controls (425.22,175.43) and (423.66,177) .. (421.72,177) .. controls (419.79,177) and (418.22,175.43) .. (418.22,173.5) -- cycle ;
		%Shape: Circle [id:dp13485030973530532] 
		\draw  [fill={rgb, 255:red, 0; green, 0; blue, 0 }  ,fill opacity=1 ] (383.72,225.5) .. controls (383.72,223.57) and (385.29,222) .. (387.22,222) .. controls (389.16,222) and (390.72,223.57) .. (390.72,225.5) .. controls (390.72,227.43) and (389.16,229) .. (387.22,229) .. controls (385.29,229) and (383.72,227.43) .. (383.72,225.5) -- cycle ;
		%Shape: Arc [id:dp9970350385929053] 
		\draw  [draw opacity=0] (386.82,228.61) .. controls (385.45,233.87) and (380.66,237.75) .. (374.97,237.75) .. controls (368.21,237.75) and (362.72,232.27) .. (362.72,225.5) .. controls (362.72,218.73) and (368.21,213.25) .. (374.97,213.25) .. controls (380.69,213.25) and (385.5,217.17) .. (386.84,222.47) -- (374.97,225.5) -- cycle ; \draw  [color={rgb, 255:red, 255; green, 0; blue, 24 }  ,draw opacity=1 ] (386.82,228.61) .. controls (385.45,233.87) and (380.66,237.75) .. (374.97,237.75) .. controls (368.21,237.75) and (362.72,232.27) .. (362.72,225.5) .. controls (362.72,218.73) and (368.21,213.25) .. (374.97,213.25) .. controls (380.69,213.25) and (385.5,217.17) .. (386.84,222.47) ;  
		%Shape: Arc [id:dp9280134663751515] 
		\draw  [draw opacity=0] (417.61,172.79) .. controls (412.87,171.1) and (409.47,166.57) .. (409.47,161.25) .. controls (409.47,154.48) and (414.96,149) .. (421.72,149) .. controls (428.49,149) and (433.97,154.48) .. (433.97,161.25) .. controls (433.97,166.85) and (430.22,171.57) .. (425.09,173.03) -- (421.72,161.25) -- cycle ; \draw  [color={rgb, 255:red, 255; green, 0; blue, 24 }  ,draw opacity=1 ] (417.61,172.79) .. controls (412.87,171.1) and (409.47,166.57) .. (409.47,161.25) .. controls (409.47,154.48) and (414.96,149) .. (421.72,149) .. controls (428.49,149) and (433.97,154.48) .. (433.97,161.25) .. controls (433.97,166.85) and (430.22,171.57) .. (425.09,173.03) ;  
		%Shape: Arc [id:dp485383204617178] 
		\draw  [draw opacity=0] (456.59,222.47) .. controls (457.95,217.03) and (462.86,213) .. (468.72,213) .. controls (475.63,213) and (481.22,218.6) .. (481.22,225.5) .. controls (481.22,232.4) and (475.63,238) .. (468.72,238) .. controls (463.09,238) and (458.33,234.28) .. (456.77,229.16) -- (468.72,225.5) -- cycle ; \draw  [color={rgb, 255:red, 255; green, 0; blue, 24 }  ,draw opacity=1 ] (456.59,222.47) .. controls (457.95,217.03) and (462.86,213) .. (468.72,213) .. controls (475.63,213) and (481.22,218.6) .. (481.22,225.5) .. controls (481.22,232.4) and (475.63,238) .. (468.72,238) .. controls (463.09,238) and (458.33,234.28) .. (456.77,229.16) ;  
		%Straight Lines [id:da007472582266312644] 
		\draw    (421.72,173.5) -- (456.22,225.5) ;
		%Shape: Circle [id:dp3020248935550284] 
		\draw  [fill={rgb, 255:red, 0; green, 0; blue, 0 }  ,fill opacity=1 ] (612.72,225.5) .. controls (612.72,223.57) and (614.29,222) .. (616.22,222) .. controls (618.16,222) and (619.72,223.57) .. (619.72,225.5) .. controls (619.72,227.43) and (618.16,229) .. (616.22,229) .. controls (614.29,229) and (612.72,227.43) .. (612.72,225.5) -- cycle ;
		%Shape: Circle [id:dp5043371844938455] 
		\draw  [fill={rgb, 255:red, 0; green, 0; blue, 0 }  ,fill opacity=1 ] (578.22,173.5) .. controls (578.22,171.57) and (579.79,170) .. (581.72,170) .. controls (583.66,170) and (585.22,171.57) .. (585.22,173.5) .. controls (585.22,175.43) and (583.66,177) .. (581.72,177) .. controls (579.79,177) and (578.22,175.43) .. (578.22,173.5) -- cycle ;
		%Shape: Circle [id:dp2228813121389871] 
		\draw  [fill={rgb, 255:red, 0; green, 0; blue, 0 }  ,fill opacity=1 ] (543.72,225.5) .. controls (543.72,223.57) and (545.29,222) .. (547.22,222) .. controls (549.16,222) and (550.72,223.57) .. (550.72,225.5) .. controls (550.72,227.43) and (549.16,229) .. (547.22,229) .. controls (545.29,229) and (543.72,227.43) .. (543.72,225.5) -- cycle ;
		%Shape: Arc [id:dp048687661248702274] 
		\draw  [draw opacity=0] (546.82,228.61) .. controls (545.45,233.87) and (540.66,237.75) .. (534.97,237.75) .. controls (528.21,237.75) and (522.72,232.27) .. (522.72,225.5) .. controls (522.72,218.73) and (528.21,213.25) .. (534.97,213.25) .. controls (540.69,213.25) and (545.5,217.17) .. (546.84,222.47) -- (534.97,225.5) -- cycle ; \draw  [color={rgb, 255:red, 255; green, 0; blue, 24 }  ,draw opacity=1 ] (546.82,228.61) .. controls (545.45,233.87) and (540.66,237.75) .. (534.97,237.75) .. controls (528.21,237.75) and (522.72,232.27) .. (522.72,225.5) .. controls (522.72,218.73) and (528.21,213.25) .. (534.97,213.25) .. controls (540.69,213.25) and (545.5,217.17) .. (546.84,222.47) ;  
		%Shape: Arc [id:dp9584165514145847] 
		\draw  [draw opacity=0] (577.61,172.79) .. controls (572.87,171.1) and (569.47,166.57) .. (569.47,161.25) .. controls (569.47,154.48) and (574.96,149) .. (581.72,149) .. controls (588.49,149) and (593.97,154.48) .. (593.97,161.25) .. controls (593.97,166.85) and (590.22,171.57) .. (585.09,173.03) -- (581.72,161.25) -- cycle ; \draw  [color={rgb, 255:red, 255; green, 0; blue, 24 }  ,draw opacity=1 ] (577.61,172.79) .. controls (572.87,171.1) and (569.47,166.57) .. (569.47,161.25) .. controls (569.47,154.48) and (574.96,149) .. (581.72,149) .. controls (588.49,149) and (593.97,154.48) .. (593.97,161.25) .. controls (593.97,166.85) and (590.22,171.57) .. (585.09,173.03) ;  
		%Shape: Arc [id:dp31497605589944344] 
		\draw  [draw opacity=0] (616.59,222.47) .. controls (617.95,217.03) and (622.86,213) .. (628.72,213) .. controls (635.63,213) and (641.22,218.6) .. (641.22,225.5) .. controls (641.22,232.4) and (635.63,238) .. (628.72,238) .. controls (623.09,238) and (618.33,234.28) .. (616.77,229.16) -- (628.72,225.5) -- cycle ; \draw  [color={rgb, 255:red, 255; green, 0; blue, 24 }  ,draw opacity=1 ] (616.59,222.47) .. controls (617.95,217.03) and (622.86,213) .. (628.72,213) .. controls (635.63,213) and (641.22,218.6) .. (641.22,225.5) .. controls (641.22,232.4) and (635.63,238) .. (628.72,238) .. controls (623.09,238) and (618.33,234.28) .. (616.77,229.16) ;  
		
		% Text Node
		\draw (272.5,38.4) node [anchor=north west][inner sep=0.75pt]  [font=\small]  {$n_{1}$};
		% Text Node
		\draw (231,97.9) node [anchor=north west][inner sep=0.75pt]  [font=\small]  {$n_{2}$};
		% Text Node
		\draw (282.5,97.9) node [anchor=north west][inner sep=0.75pt]  [font=\small]  {$n_{3}$};
		% Text Node
		\draw (179,57.9) node [anchor=north west][inner sep=0.75pt]    {$+$};
		% Text Node
		\draw (152.5,290.4) node [anchor=north west][inner sep=0.75pt]  [font=\small]  {$n$};
		% Text Node
		\draw (200,286.4) node [anchor=north west][inner sep=0.75pt]    {$=Z_{n}^{( 1)}$};
		% Text Node
		\draw (318.5,301.4) node [anchor=north west][inner sep=0.75pt]  [font=\small]  {$n_{i}$};
		% Text Node
		\draw (391,301.9) node [anchor=north west][inner sep=0.75pt]  [font=\small]  {$n_{j}$};
		% Text Node
		\draw (413,284.4) node [anchor=north west][inner sep=0.75pt]    {$=\tilde{\mathsf{g}}_{n_{i} ,n_{j}}( z_{i} ,z_{j})$};
		% Text Node
		\draw (108,37.9) node [anchor=north west][inner sep=0.75pt]  [font=\small]  {$n_{1}$};
		% Text Node
		\draw (65,97.9) node [anchor=north west][inner sep=0.75pt]  [font=\small]  {$n_{2}$};
		% Text Node
		\draw (116.5,97.9) node [anchor=north west][inner sep=0.75pt]  [font=\small]  {$n_{3}$};
		% Text Node
		\draw (430.5,38.4) node [anchor=north west][inner sep=0.75pt]  [font=\small]  {$n_{1}$};
		% Text Node
		\draw (389,97.9) node [anchor=north west][inner sep=0.75pt]  [font=\small]  {$n_{2}$};
		% Text Node
		\draw (440.5,97.9) node [anchor=north west][inner sep=0.75pt]  [font=\small]  {$n_{3}$};
		% Text Node
		\draw (337,57.9) node [anchor=north west][inner sep=0.75pt]    {$+$};
		% Text Node
		\draw (590.5,38.4) node [anchor=north west][inner sep=0.75pt]  [font=\small]  {$n_{1}$};
		% Text Node
		\draw (549,97.9) node [anchor=north west][inner sep=0.75pt]  [font=\small]  {$n_{2}$};
		% Text Node
		\draw (600.5,97.9) node [anchor=north west][inner sep=0.75pt]  [font=\small]  {$n_{3}$};
		% Text Node
		\draw (497,57.9) node [anchor=north west][inner sep=0.75pt]    {$+$};
		% Text Node
		\draw (272.72,169.4) node [anchor=north west][inner sep=0.75pt]  [font=\small]  {$n_{1}$};
		% Text Node
		\draw (231.22,228.9) node [anchor=north west][inner sep=0.75pt]  [font=\small]  {$n_{2}$};
		% Text Node
		\draw (282.72,228.9) node [anchor=north west][inner sep=0.75pt]  [font=\small]  {$n_{3}$};
		% Text Node
		\draw (179.22,188.9) node [anchor=north west][inner sep=0.75pt]    {$+$};
		% Text Node
		\draw (108.22,168.9) node [anchor=north west][inner sep=0.75pt]  [font=\small]  {$n_{1}$};
		% Text Node
		\draw (65.22,228.9) node [anchor=north west][inner sep=0.75pt]  [font=\small]  {$n_{2}$};
		% Text Node
		\draw (116.72,228.9) node [anchor=north west][inner sep=0.75pt]  [font=\small]  {$n_{3}$};
		% Text Node
		\draw (430.72,169.4) node [anchor=north west][inner sep=0.75pt]  [font=\small]  {$n_{1}$};
		% Text Node
		\draw (389.22,228.9) node [anchor=north west][inner sep=0.75pt]  [font=\small]  {$n_{2}$};
		% Text Node
		\draw (440.72,228.9) node [anchor=north west][inner sep=0.75pt]  [font=\small]  {$n_{3}$};
		% Text Node
		\draw (337.22,188.9) node [anchor=north west][inner sep=0.75pt]    {$+$};
		% Text Node
		\draw (590.72,169.4) node [anchor=north west][inner sep=0.75pt]  [font=\small]  {$n_{1}$};
		% Text Node
		\draw (549.22,228.9) node [anchor=north west][inner sep=0.75pt]  [font=\small]  {$n_{2}$};
		% Text Node
		\draw (600.72,228.9) node [anchor=north west][inner sep=0.75pt]  [font=\small]  {$n_{3}$};
		% Text Node
		\draw (497.22,188.9) node [anchor=north west][inner sep=0.75pt]    {$+$};
		% Text Node
		\draw (11,187.9) node [anchor=north west][inner sep=0.75pt]    {$+$};
		\end{tikzpicture}	
	\caption{The diagrams that constitute the $N=3$ contribution to the partition function 
		and are the one-to-one correspondence with the contributions to the representation \eqref{cIfromMQM-set} of $\cIi{3}_{\bfn}$.
		The connected graphs in the first line correspond to contributions with $m=1$ and 
		give the $N=3$ term to the free energy \eqref{Fnp-int}. The next three terms correspond to contributions with $m=2$ 
		and the last term provides the $m=3$ contribution.}
	\label{fig-graphs}
	\vspace{-0.6cm}
\end{figure}

Having identified the partition functions, one can write an expression for the free energy 
in terms of string amplitudes. 
Of course, to this end, one could simply take the logarithm of the partition function \eqref{Znp-res}.
However, one can do better by noticing that the expression \eqref{encFnpZZ}
can be rewritten through the functions \eqref{def-Lami}
\be
\cF_{\rm np,D}(\mu)= \sum_{N=1}^\infty \frac{1}{N}\prod_{i=1}^N \[\sum_{n_i=1}^\infty \Zi{1}_{n_i} \]\Lami{N}_\bfn,
\label{Fnp-Lam}
\ee
where the factors $\Zi{1}_n$ have been defined in \eqref{normfac}.
Next, one observes that the factor $(N-1)!\Lami{N}_\bfn$ is the $m=1$ term in the representation \eqref{cIfromMQM-set}
of the integral $\cIi{N}_{\bfn}$. From its derivation, it should be clear that 
this term corresponds to contributions generated by residues at $z_i=z_j/(\lambda_i\lambda_j)$ for all $i>1$.
On the other hand, if at least for one $z_i$ with $i>1$ we pick up the residue at $z_i=0$, 
one obtains a term contributing to the part of \eqref{cIfromMQM-set} with $m>1$.
Thus, one can extract the $m=1$ term if one subtracts all contributions from residues at $z_i=0$.
Naively, it is sufficient to replace $\gsf_{n_i,n_j}(z_i,z_j)$ \eqref{def-sfg} by 
\be 
\tgsf_{n_i,n_j}(z_i,z_j)=\gsf_{n_i,n_j}(z_i,z_j)-1
=\frac{(\lambda_i^2-1)(\lambda_j^2-1) z_i z_j}{(z_i-\lambda_i\lambda_j z_j)(\lambda_i\lambda_j z_i-z_j)}\, .
\label{def-sftg}
\ee 
However, since for $N>2$ there are more factors of $\gsf_{n_i,n_j}$ than the residues to be evaluated, 
this replacement would generate too many subtractions. In fact, it is easy to realize that the terms to be left
can all be written as products of $\tgsf_{n_i,n_j}(z_i,z_j)$ such that the pairs $(ij)$ ``connect'' all indices from 1 to $N$.
As a result, we arrive at the following representation
\be
\cF_{\rm np,D}(\mu)= \sum_{N=1}^\infty \frac{1}{N!}\prod_{i=1}^N \[\sum_{n_i=1}^\infty \Zi{1}_{n_i} \]\
\tcIi{N}_\bfn,
\label{Fnp-int}
\ee
where
\be  
\tcIi{N}_\bfn=\oint \prod_{i=1}^N \frac{d z_i}{2\pi i z_i} \, \sum_{\gamma\in \Gamma_N} \prod_{(ij)\in \gamma} \tgsf_{n_i,n_j}(z_i,z_j),
\label{deftInz}
\ee
$\gamma_N$ is the set of {\it connected} graphs with $N$ vertices, 
and $(ij)$ denotes an edge connecting the vertices labeled by $i$ and $j$. In Fig. \ref{fig-graphs},
we illustrate the relation between graphs and contributions to the partition function \eqref{Znp-res} for $N=3$.

Remarkably, this result can be identified with the D-instanton induced effective action 
emerging from string field theory, which has recently been derived in \cite{Sen:2024zqr}. 
The $N$-th instanton term in this effective action is given by $1/N!$ multiplying the integral over instanton zero modes
of the so called {\it target space connected contributions}. They include contributions from all connected 
and disconnected worldsheets that end on D-instantons in such a way that the whole surface in the target space is connected.
This precisely corresponds to the formula \eqref{Fnp-int}, with the additional restriction to only disk and annulus amplitudes.

More concretely, let $\whS'_{i_1\cdots i_r}$ denotes the contribution of all connected worldsheets with boundaries 
lying on one of the D-instantons labeled by $i_s$, $s=1,\dots ,r$, with at least one boundary on each of them.
Then, if we identify 
\be 
e^{\whS'_i}=\Zi{1}_{n_i},
\qquad
e^{\whS'_{ij}}=\gsf_{n_i,n_j}(z_i,z_j)
\label{ident-hS}
\ee 
and set $\whS'_{ijk}=0$, the first three terms in \eqref{Fnp-int} reproduce (2.6) and (3.7) in \cite{Sen:2024zqr},
up to a factor $\cN$ that can be set to 1 in our context.
Furthermore, these identifications imply that 
\be
e^{S_{(r)}-S_{(0)}}:=\int \exp\[\sum_{s=1}^r \sum_{1\leq i_1<\cdots<i_s\leq r} \whS'_{i_1\cdots i_s} \]
=\prod_{i=1}^N \[
\sum_{n_i=1}^\infty \Zi{1}_{n_i} \]\cIi{N}_\bfn,
\label{ident-SI}
\ee 
where we set to zero all $\whS'_{i_1\cdots i_r}$ with more than two indices.
This ensures that the representation for the effective action found in \cite[Eq.(3.1)]{Sen:2024zqr}
is identical to the logarithm of the partition function \eqref{Znp-res}.

Making these identifications, we neglected worldsheet topologies with negative Euler number,
which corresponds to our approximation in \eqref{Znp}. On the other hand, working within this approximation, 
we have reproduced the non-perturbative part of the free energy \eqref{encFnp} which, as follows from MQM, must be {\it exact}.
This suggests that contributions of all topologies that have been neglected actually cancel in 
2d string theory in the linear dilaton background, so that \eqref{ident-SI} is an exact equality. 
A more refined statement would require the equalities \eqref{ident-hS} and the vanishing of all $\whS'_{i_1\cdots i_r}$ with $s>2$.
It would be interesting to verify any of these matrix model predictions explicitly.

\subsection{Decompactification limit}
\label{sec-limit}

Finally, let us briefly analyze the decompactification limit of the above results.
It is easier to do this at the level of the free energy:
one should simply divide it by the length of the compact circle and send $R\to \infty$. For the non-perturbative part 
\eqref{encFnp}, this gives  
\be
\lim\limits_{R\to \infty}\frac{\cF_{\rm np}(\mu)}{2\pi R}=
\sum_{n=1}^\infty {(-1)^{n}\, e^{-2\pi n \mu}\over 8i \pi^2 n^2 }=\frac{1}{8i\pi^2}\, \Li_2\(-e^{-2\pi \mu}\).
\label{Fnp-decomp}
\ee

Of course, a similar limit can be taken for the representation \eqref{Fnp-Lam}.
It is natural to represent the result as 
\be
\lim\limits_{R\to \infty}\frac{\cF_{\rm np}(\mu)}{2\pi R}=
\sum_{N=1}^\infty \frac{1}{N!}\prod_{i=1}^N \[\sum_{n_i=1}^\infty \whZi{1}_{n_i} \]\whcIi{N}_\bfn,
\label{Fnp-int-decomp}
\ee
where 
\bea
\whZi{1}_n&=&\lim\limits_{R\to \infty}\frac{\Zi{1}_n}{2\pi R}=\frac{(-1)^{n} \zeta_{n}}{4i\pi^2 n}\, e^{-n/g_s},
\label{def-whZ1}
\\
\whcIi{N}_\bfn&=&\lim\limits_{R\to \infty} (2\pi R)^{N-1}\tcIi{N}_\bfn
=\int\limits_{-\infty}^\infty\prod_{i=1}^{N-1} dx_i\sum_{\gamma\in \Gamma_N} \prod_{(ij)\in \gamma} 
\frac{4\pi^2 n_i n_j}{x_{ij}^2-\pi^2(n_i+n_j)^2}
\nn\\
&=&(N-1)!\,(-4i\pi^2)^{N-1}\, \frac{ \prod_{i=1}^N n_i}{\sum_{i=1}^N n_i}
\label{def-whcJ}
\eea 
and, as usual, the integrals are supposed to be regularized by the Lorentzian prescription.
Note that in \eqref{def-whcJ} there is no integral over $x_N$ and it can be completely absorbed into 
a redefinition of other integration variables since the integrand depends only on the differences $x_{ij}$.
The disappearing integral is exactly the same as the one that produces the momentum conserving delta function 
in scattering amplitudes in the presence of D-instantons in the non-compact case \cite{Sen:2020cef}.
The agreement between \eqref{Fnp-decomp} and \eqref{Fnp-int-decomp} is ensured by the same identity $c_n=-\frac{1}{2n}$,
with $c_n$ defined in \eqref{res-cn}, as in the compact case.

The last remark is that, upon substitution of \eqref{reszeta}, the prefactor in \eqref{def-whZ1}
gives $-i \cN_n$ with 
\be  
\cN_n=\frac{(-1)^n}{4\pi^2 n}\, \frac{(2n-1)!!}{(2n)!!}\, .
\label{cN-BRY}
\ee 
The normalization coefficient \eqref{cN-BRY} reproduces the number found in \cite[Eq.(4.13)]{Balthazar:2019ypi}
by comparing the MQM S-matrix with the string amplitudes supplemented by the Lorentzian prescription,
while the factor $-i$ was shown to come from the Wick rotation in \cite{Sen:2020ruy}.

\section{Discussion}
\label{sec-disc}

In this paper, we have fixed the precise numerical normalization of the D-instanton string amplitudes in 2d string theory
using a correspondence, supported by resurgence theory, 
between $(1,n)$ ZZ-instantons and saddle points of an integral representation of the scattering phase in the dual MQM. 
We have shown that the resulting values of the normalization coefficients correspond 
to the Lorentzian prescription making multi-instanton string amplitudes well-defined.
We have proven this by explicitly evaluating all these amplitudes in the theory compactified on a circle of finite radius
and comparing with the known expression for the non-perturbative free energy derived in MQM.

We have also observed that the free energy expressed through string amplitudes has the same form as 
the D-instanton induced effective action in SFT. Furthermore, this identification suggests the vanishing of
contributions from worldsheet topologies with negative Euler number.
This is a very strong prediction, which certainly deserves a verification.
If it is true, there should exist some fundamental reason for such a crucial simplification,
which might also have implications for string amplitudes of higher topology in other string theories. 

Note that we have found that some previous extensions of the chiral formalism of MQM to the non-perturbative level 
were somewhat naive. 
They ignored certain subtleties explained in appendix \ref{ap-signs}, which are absent at the perturbative level.
Their correct treatment leads to various sign flips. In particular, the non-perturbative part of the free energy \eqref{encFnp}
has the opposite sign compared to \cite{Alexandrov:2004cg,Alexandrov:2023fvb}. This sign is crucial to 
have agreement with the multi-instanton string amplitudes in the Lorentzian prescription.
If we took the wrong sign, it would lead to the non-physical ``anti-Lorentzian'' prescription. 

Finally, in appendix \ref{ap-unitary}, we have analyzed consequences of accepting the unitary prescription
for multi-instanton string amplitudes. Using their explicit evaluation, we have derived the corresponding values of 
the normalization coefficients. However, we could not find a reasonable way to reproduce them from the MQM
scattering phase. 

It is natural to assume that the same values of the normalization coefficients that we derived here in the linear dilaton 
background hold also for time-dependent backgrounds obtained by introducing a tachyon perturbation.
The crucial difference of these backgrounds is that they remove the degeneracy between different types of ZZ-instantons.
On the other hand, in the MQM formalism, they preserve integrability and allow for an exact evaluation of
the instanton effects. It would be interesting to see whether the knowledge of the exact normalization
helps in finding the exact non-perturbative free energy in perturbed backgrounds.

An important open issue, which we remained unsolved, is to rederive the normalization coefficients from SFT.
They are expected to arise from Picard-Lefschetz theory applied to an integral over 
massless and tachyonic open string modes in a D-instanton configuration.
However, to perform such an analysis
would likely require to improve the knowledge of the tachyon effective potential,
which currently remains quite limited. 

In this respect it would also be interesting to understand whether negative tension branes, argued to exist and 
to be a necessary ingredient for the complete resurgent structure of string theory or its matrix model dual 
\cite{Marino:2022rpz,Schiappa:2023ned}, could play any role in the emergence of the rational normalization factors.

\bigskip 

{\bf Acknowledgments.}
We would like to thank Marcos Mari\~no and Ashoke Sen for valuable discussions.
SA is grateful to the organizers of the ``ReNewQuantum International Conference 2025 --- Resurgence in Mathematics and Physics''
in Odense, where the results of this work have been presented, for the kind hospitality. RK is supported by the Department of Atomic
Energy, Government of India, under project no. RTI4001.

\appendix 

\section{Subtleties of MQM at non-perturbative level}
\label{ap-signs}

In the main text, we computed the scattering phase $\phi_0(E)$ by imposing the relation \eqref{ennormcond1}.
It differs from the one that was imposed in \cite{Alexandrov:2004cg,Alexandrov:2023fvb}
and had the form
\be
\hat S' [\psep](\xm)\equiv
\frac{1}{ \sqrt{2\pi}}\int_0^\infty d\xp \, e^{ i\xp\xm} \psep(\xp)=\psem(\xm).
\label{Spm}
\ee
One could think that $\hat S'$ is the inverse of $\hat S$ and therefore the two relations 
are equivalent. However, this is not the case, which is, in particular, manifested by the fact that
they lead to {\it complex conjugate} expressions for the scattering phase $\phi_0(E)$.
The reason is very simple: the two operators were inverse of each other only if we integrated over the whole real line. 
But then we would have to include another set of eigenfunctions corresponding to fermions on the ``other side of the potential'',
which is equivalent to considering ``the theory of type II'' in the terminology of \cite{Moore:1991zv,Alexandrov:2002fh}.
Alternatively, one could take the integration kernel to be $2\cos(\xp\xm)$ instead of $e^{ i\xp\xm}$, 
as was originally done in \cite{Alexandrov:2002fh}. But then the S-matrix would be unitary and 
non-perturbative effects would be absent.

Thus, we are in a situation where there are two possible ways to fix the scattering phase and 
we should find a way to determine which of them is the physically correct one.
To this end, we suggest the following reasoning. One expects that the scattering coefficient arises 
by applying the S-matrix to an incoming wave function and expressing it through outgoing wave functions.
Thus, to choose between $\hat S$ and $\hat S'$, one should understand which of the chiral representations
describes incoming and which outgoing states.
It is easy to see that any function $\Psi_\pm(\xpm)$ can be promoted to a solution of the time-dependent 
Schr\"odinger equation with the Hamiltonian \eqref{chiralH} by setting $\Psi_\pm(\xpm,t)=e^{\mp t/2}\Psi_\pm( e^{\mp t}\xpm)$.
Hence, for $\Psi_\pm(\xpm)$ describing a bump at finite values of the argument,
the bump will move to large $\xpm$ at $t\to \pm\infty$, respectively. 
This implies that the $\xm$-representation should be considered as describing incoming states, 
and the $\xp$-representation as describing outgoing states. In turn, this favors $\hat S$ \eqref{ennormcond1}
as the correct S-matrix operator.
This result is in agreement with \cite{Balthazar:2019rnh,Balthazar:2019ypi} where the reflection coefficient 
coincides with the one computed in \eqref{scatphase0}.

The flip of the sign of $\Im\phi_0(E)$ compared to \cite{Alexandrov:2004cg,Alexandrov:2023fvb}
has important implications for the non-perturbative part of the free energy. 
Due to the relation \eqref{enepart}, it must also flip the sign, which leads us to the result \eqref{encFnp}. 
For completeness, we also provide an integral representation for the free energy in the linear dilaton background, 
which extends a similar integral representation from \cite{Klebanov:1991qa} to the non-perturbative level and 
corrects such an extension given in \cite{Alexandrov:2004cg}:
\be  
\cF(\mu)=-\frac14\, \int_{\Lambda^{-1}}^\infty \frac{ds}{s}\, \frac{e^{-i\mu s}}{\sinh\frac{s}{2}\sinh\frac{s}{2R}}\, ,
\ee 
where $\Lambda$ is a cut-off affecting only non-universal terms.

It is worth noting that if we stick to the Euclidean theory and do not try to connect to 
its Lorentzian version, the above reasoning cannot be applied, and both signs of the non-perturbative effects
seem to be allowed. This corresponds to the fact that, from the pure Euclidean viewpoint,
there is no way to choose between Lorentzian and ``anti-Lorentzian'' prescription for the contour
in the integral representation of amplitudes.
Thus, it is the connection to the Lorentzian version of the theory that provides the physical input
allowing us to resolve the ambiguity.

\section{Unitary prescription}
\label{ap-unitary}

Let us assume that it is the unitary prescription of \cite{Sen:2020ruy} that is used to regularize
multi-instanton string amplitudes and compute the multipliers $\chzeta_n$ corresponding to this choice.
Here and below we use the check mark $\check \cdot$ to distinguish quantities computed with this prescription.

As explained in section \ref{subsec-prescrip}, the unitary prescription is obtained by 
averaging over the Lorentzian prescription and its complex conjugate.
Since the shift by $i\eps$ introduced by the Lorentzian prescription is the only source of complexity 
in the integrals $\cIi{N}_\bfn$ \eqref{defIn}, their values under the unitary prescription 
are given simply by the real part of the expression \eqref{cIfromMQM}.
Performing the manipulations done in section \ref{subsec-disentangle} backwards, one finds that 
this amounts to dropping terms in $\Ci{N}_\bfn$ \eqref{def-Cbfn} with $N-m$ odd 
and in $c_n$ \eqref{res-cn} with $k$ even.\footnote{The latter fact is easier to see by noticing that 
	in the unitary prescription the integrals $\tcIi{N}_\bfn$ remain the same for $N$ odd and vanish for $N$ even,
	and then comparing \eqref{Fnp-int} with \eqref{encFnpZZ} and \eqref{res-cn}.}
Thus, we have 
\be  
c_n=\sum_{k=1}^n \frac{(-1)^{k}-1}{2k}\sum_{\sum_{j=1}^k n_j=n} \prod_{j=1}^k\chzeta_{n_j}\, .
\label{res-cn-new}
\ee 
Recombining these numbers into a generating series, one finds
\be 
\sum_{n=1}^\infty c_n x^n=\sum_{k=1}^\infty  \frac{(-1)^{k}-1}{2k}\prod_{i=1}^k \sum_{n_i=1}^\infty \chzeta_{n_i}x^{n_i}
=\hf\,\log \frac{2-\chf(x)}{\chf(x)}\, ,
\label{gencn}
\ee 
where we introduced the generating series of the new multipliers $\chf(x)=\sum_{n=0}^\infty \chzeta_n x^n$.
On the other hand, $c_n$ should still be equal to $-\frac{1}{2n}$ to reproduce the free energy \eqref{encFnp},
which implies that the generating series \eqref{gencn} should be equal to $\hf\log(1-x)$.
Solving for $\chf(x)$, one finds
\be 
\chf(x)=\frac{1}{1-x/2}\ \Rightarrow \ \chzeta_n=2^{-n}.
\label{val-chzeta}
\ee 
This result agrees with \cite{Sen:2020ruy} where $\chzeta_n$ 
(more precisely, the analogue of the normalization coefficients \eqref{cN-BRY})
have been computed up to $n=3$.

However, the function $\chf(-e^{-2\pi\mu})=\(1+\hf\, e^{-2\pi\mu}\)^{-1}$ does not appear naturally in MQM.
For instance, because of the coefficient $1/2$, it is not related to the Stokes factor of the Gamma function.
One can try to reconstruct the scattering phase of the matrix model corresponding to this function by requiring that
\be
\Im \check\phi_0(E)=\log\chf(-e^{2\pi E}) 
=-\log\(1+\hf\, e^{2\pi E}\).
\label{newIm-phase}
\ee
If one leaves the real part of the scattering phase unchanged, this gives
\be
e^{-i\check\phi_0(E)}=\frac{e^{\frac{-\pi i}{4}-\frac{\pi}{2}\, E}}{\sqrt{2\pi}}\,
\frac{\(1+e^{2\pi E}\)^{1/2}}{1+\hf\, e^{2\pi E}}\,\Gamma\(\hf-iE\).
\label{scatphase-new}
\ee
Such a scattering phase is not expected to arise in any reasonable version of the theory.
Alternatively, one can drop our main assumption that the saddle points of the integral representation 
of the scattering phase \eqref{phase0} correspond to $(1,n)$ ZZ-instantons, but keep 
its imaginary part \eqref{Im-scatphase} intact. Then it can be expressed through the multipliers $\chzeta_n$ as follows 
\be 
e^{\Im\phi_0(E)}=\(\frac{1+\sum_{n=1}^\infty \chzeta_n (-1)^n e^{2\pi E}}{1-\sum_{n=1}^\infty \chzeta_n (-1)^n e^{2\pi E}}\)^{1/2}.
\ee 
This expression allows to read off the contribution of each $(1,n)$ ZZ-brane into the $n$-th saddle point
in this modified prescription. But it is not clear to us how this complicated pattern can be justified.
This is to be compared with the simplicity of our original proposal, which automatically leads to the Lorentzian prescription.
Thus, we conclude that, even if the unitary prescription reproduces the non-perturbative free energy 
by properly adjusting the multipliers $\zeta_n$, it seems to be incompatible with any natural identification
between ZZ-branes and saddle points of the matrix model integral.

\providecommand{\href}[2]{#2}\begingroup\raggedright\endgroup

%\bibliography{combined}
%\bibliographystyle{utphys}

\end{document}